\documentclass[jkps,twoside,twocolumn]{revtex4}
\usepackage{amssymb}
\usepackage{amsmath}
\usepackage{graphicx}
\usepackage{color}

\newcommand{\be}{\begin{eqnarray}}
\newcommand{\ee}{\end{eqnarray}}
\newcommand{\bbm}{\begin{bmatrix}}
\newcommand{\ebm}{\end{bmatrix}}
\newcommand{\A}{\mathrm{A}}
\newcommand{\B}{\mathrm{B}}

\newcommand{\pr}{\prime}

\begin{document}
\title{Self-similar occurrence of massless Dirac particles in graphene under magnetic field}

\author{Jun-Won \surname{Rhim}}
\affiliation{School of Physics, Korea Institute for Advanced Study, Seoul 130-722, Korea }
\author{Kwon \surname{Park}}
\affiliation{School of Physics, Korea Institute for Advanced Study, Seoul 130-722, Korea }
\date{\today}

\begin{abstract}
Intricate interplay between the periodicity of the lattice structure and that of the cyclotron motion gives rise to a well-known self-similar fractal structure of the energy eigenvalue, known as the Hofstadter butterfly, for an electron moving in lattice under magnetic field.
Evolving from the $n=0$ Landau level, the central band of the Hofstadter butterfly is especially interesting since it may hold a key to the mysteries of the fractional quantum Hall effect observed in graphene.
While the entire Hofstadter butterfly can be in principle obtained by solving Harper's equations numerically, the weak-field limit, most relevant for experiment, is intractable due to the fact that the size of the Hamiltonian matrix, that needs to be diagonalized, diverges.
In this paper, we develop an effective Hamiltonian method that can be used to provide an accurate analytic description of the central Hofstadter band in the weak-field regime.
One of the most important discoveries obtained in this work is that massless Dirac particles always exist inside the central Hofstadter band no matter how small the magnetic flux may become.
In other words, with its bandwidth broadened by the lattice effect, the $n=0$ Landau level contains massless Dirac particles within itself.  
In fact, by carefully analyzing the self-similar recursive pattern of the central Hofstadter band, we conclude that massless Dirac particles should occur under arbitrary magnetic field.
As a corollary, the central Hofstadter band also contains a self-similar structure of recursive Landau levels associated with such massless Dirac particles. 
To assess the experimental feasibility of observing massless Dirac particles inside the central Hofstadter band, we compute the width of the central Hofstadter band as a function of magnetic field in the weak-field regime.
\end{abstract}



\maketitle

\section{Introduction}

Observing the behavior of electrons in graphene under high magnetic field has played an important role not only for uncovering new quantum Hall states, but also for proving the very existence of  massless Dirac particles~\cite{Novoselov,PKim1}.
Affected by the linear dispersion near Dirac points, Landau levels are formed in graphene such that their energy is scaled as $\textrm{sgn}(n)\sqrt{|n|}$ in units of $\sqrt{2} \hbar v_F/l_B$ with $n$, the Landau level index, allowed for all integers including positive, zero, and negative~\cite{Ando}.
In the above, $v_F$ is the Fermi velocity at the Dirac point and $l_B=\sqrt{\hbar c/eB}$ is the magnetic length.
The $n=0$ Landau level offers a particularly intriguing departure from the usual quantum Hall effect (QHE) in that its Hall coefficient is shifted by half an integer.
With both spin and valley degeneracy taken into account, the consequent Hall conductance is predicted to be quantized in the form of $4(n+1/2)$  in units of $e^2/h$, which exhibits beautiful agreement with experiment~\cite{Novoselov,PKim1}.

There is, however, a glaring omission in the discussion so far.
In the above, the effect of lattice is completely ignored except that the electron dispersion becomes linear near Dirac points.
The question is how valid this assumption can be.
More specifically, will there be any changes in the Landau-level structure once the effect of lattice is better incorporated?
Na\"{i}vely speaking, since the deviation from the linear dispersion occurs in relatively high energy, one may expect that the Landau levels should be more or less the same as before so that they remain as flat bands.
In particular, the $n=0$ Landau level is then expected to remain as a flat band pinned exactly at zero energy due to the particle-hole symmetry.
Seemingly innocuous, if true, this expectation gives rise to a very puzzling question:
what determines which states within the $n=0$ Landau level evolves into the particle (or the positive energy) branch and which into the hole (or the negative energy) branch at the edge?
A natural resolution of this puzzle is that the $n=0$ Landau level is broadened
with its bandwidth becoming finite.
If so, what would be the nature of such bandwidth-broadened $n=0$ Landau level?

The quantum mechanical problem of an electron moving in lattice under magnetic field is generally known as the Azbel-Hofstadter problem named after Azbel~\cite{Azbel}, who originally proposed the model, and Hofstadter~\cite{Hofstadter}, who first obtained a numerical solution in the square lattice and showed the existence of a self-similar fractal structure in energy eigenvalue, dubbed as the Hofstadter butterfly.
The actual equations, that need to be solved, are known as Harper's equations which are in fact nothing but the energy eigenvalue equation for the Hamiltonian matrix.    
By numerically solving Harper's equations, the self-similar fractal structure of the Azbel-Hofstadter model was found also for various other lattices including the triangular and the honeycomb lattice~\cite{Claro,Rammal,Hasegawa,Hatsugai1}.

In addition to numerical studies solving Harper's equation, there have been extensive efforts to obtain analytic solutions~\cite{Rauh,Kohmoto1,Freed,Rammal2,Zak,Barelli,Gedik,Krasovsky,Wiegmann,Hatsugai,Faddeev,Hatsugai2,Abanov,Krasovsky2,Hoshi,Kohmoto2,Delplace}.
The reason for such efforts is multifaceted. 
For one, many researchers have been curious about the very origin of the self-similar fractal structure seen in the Hofstadter butterfly and tried to make a connection to other known systems exhibiting similar fractal structures.
For another, numerical computations can be performed only in the situation where the magnetic flux per unit cell, $\phi$, is a rational fraction of the magnetic flux quantum, $\phi_0=hc/e$. 
Therefore, what happens at irrational fractions can be addressed only by the analytic approaches.
Perhaps, the most important reason in connection with experiment is the fact that the numerical approach cannot access the weak-field limit where the size of the matrix that needs to be diagonalized diverges.
The weak-field limit is most relevant for experiment since, even in the quantum Hall regime, the magnetic flux per unit cell is typically much less than $1/100$ in units of magnetic flux quantum.

Among various analytic approaches, the Bethe-ansatz approach is regarded to be most systematic, where the Azbel-Hofstader problem is converted into solving the Bethe-ansatz equations whose roots are directly connected to the energy eigenvalues as well as eigenstates.
Despite providing such insightful relationship to an integrable model, the Bethe-ansatz approach is proven to be of little practical use since the Bethe-ansatz equations are generally insoluble except for special cases.
The use of other analytic approaches is also similarly limited.

In this paper, we develop a method that can be used to provide an accurate analytic description of the evolution of the $n=0$ Landau level as a function of magnetic field ranging from being arbitrarily weak to moderately strong.
In this method, it is shown that, for $\phi/\phi_0=p/q$ with $p$ and $q$ being coprime positive integers, the central band of the Hofstadter butterfly, which is obtained from the original $2q \times 2q$ matrix for Harper's equations, is captured extremely accurately by diagonalizing the effective Hamiltonian matrix with a much reduced size of $2p \times 2p$ in the weak-field regime.
The central band of the Hofstadter butterfly is connected with the $n=0$ Landau level in the continuum limit.
Actually, this effective Hamiltonian matrix works quite well for $\phi/\phi_0$ as large as $0.3$.
One of the most important discoveries of this work is that, no matter how small the magnetic flux per unit cell may become, the central Hofstadter band (CHB) always contains massless Dirac particles whose energy dispersion is completely isomorphic to that in the absence of magnetic field.
In fact, by combining the self-similar pattern of the central Hofstadter band and some analytic as well as numerical results for the zero-energy modes of Harper's equations, we conclude that there should be exactly $2q$ Dirac cones in the magnetic Brillouin zone (MBZ)  for $\phi/\phi_0=p/q$ with arbitrary $p$ and $q$.
A corollary of this result is that there should also be a self-similar occurrence of Landau levels associated with such Dirac cones.

In order to assess the experimental feasibility of observing such massless Dirac particles within the central Hofstadter band, we compute the width of the central Hofstadter band which, for small $\phi/\phi_0$, is predicted to scale as $\exp{(-\gamma \frac{\phi_0}{\phi})}$ in units of the energy level spacing between the $n=0$ and $1$ Landau level, $\sqrt{2} \hbar v_F/l_B$.
Here, $\gamma=|\textrm{Cl}_2(5\pi/3)|/\pi \simeq 0.323$ and $\textrm{Cl}_2 (\theta) = \sum^{\infty}_{n=1} \sin{(n\theta)}/n^2$ is called the Clausen function.
Actually, motivated by an intriguing conjecture proposed by Thouless~\cite{Thouless0} a while ago, there has been a long history for addressing how the total bandwidth of the Hofstadter butterfly scales as a function of magnetic field~\cite{Thouless1,Thouless2,Watson,Helffer,Krasovsky,Rammal,Ketzmerick}.
To the best of our knowledge, our result is the first report for the scaling of the Hofstadter butterfly bandwidth in the honeycomb lattice.
Considering difficulties in directly observing the Hofstadter butterfly under magnetic field with typically available strength, we believe that a precise measurement of the bandwidth itself can be used to infer the existence of the Hofstadter butterfly in addition to the Diophantine equation for the quantized Hall conductance~\cite{Thouless3,Streda,Klitzing,Macdonald,Hasegawa2}.

The rest of the paper is organized as follows.
In Sec.~\ref{section:model}, we present the Azbel-Hofstadter model in graphene with a particular choice of gauge called the optimal gauge.
In Sec.~\ref{section:zero-energy}, we analyze various properties of the zero-energy solutions for Harper's equations, which play a crucial role in our effective Hamiltonian method by generating basis wave functions for the central Hofstadter band.
A precise mathematical form of the effective Hamiltonian is presented in Sec.~\ref{section:effective_Hamiltonian}, where it is shown that the resulting magnetic band structure provides an excellent agreement with that of the central Hofstadter band obtained from the original Harper's equations in the weak-field regime.
In Sec.~\ref{section:self-similar}, by using such effective Hamiltonian method, we carefully analyze the self-similar recursive pattern of the central Hofstadter band, which is then combined with analytic as well as numerical results for the zero-energy modes to show that massless Dirac particles should occur under arbitrary magnetic field.
We conclude in Sec.~\ref{section:conclusion}.

\section{Azbel-Hofstadter problem for graphene}
\label{section:model}

\begin{figure}
\includegraphics[width=1\columnwidth]{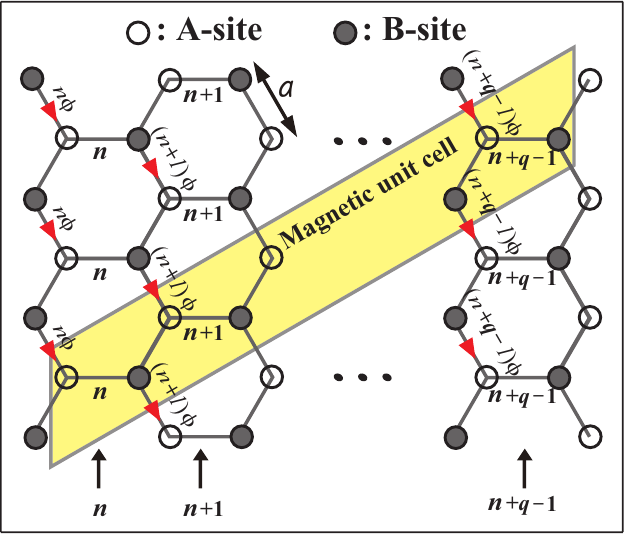}
\caption{(Color online) Schematic diagram for the gauge used in this work.
The yellow parallelogram depicts a magnetic unit cell (MUC).
Magnetic unit cells are denoted by the MUC index, $\alpha$, along the $y$-direction.
Different carbon sites within the same magnetic unit cell are distinguished by the dimer index, $n$, and the A/B sublattice index.
Note that horizontally-connected A and B carbon sites share the same dimer index.
Red arrows indicate the directions of the paths, along which non-zero phases are gained via the Peierls substitution. 
The value of the non-zero Peierls phase is written near each arrow while all the other phases are zero.
This gauge is called the optimal gauge.}
\label{fig:optimal_gauge}
\end{figure}

The Azbel-Hofstadter problem is nothing but an energy eigenvalue problem of the tight-binding Hamiltonian under magnetic field:
\begin{equation}
\mathbf{H}=\sum_{\langle i,j \rangle} t_{ij} c^\dagger_i c_j \;,
\end{equation}
where $t_{ij}$ is the hopping amplitude between the nearest-neighboring sites with its phase determined via the Peierls substitution, $t_{ij} = t_0 \rightarrow t_0 e^{2\pi\phi_{ij}}$, where $\phi_{ij}=\frac{e}{2\pi\hbar c} \int^j_i {\bf A}\cdot d{\bf l}$ and ${\bf A}$ is the vector potential.
Here, $t_0$ is the hopping amplitude in the absence of external magnetic field.
For convenience, we now fix the energy scale by setting $t_0=1$.
The physical energy scale can be restored by re-introducing $t_0$, when necessary.
While any vector potential satisfying the condition that the contour integral, $\oint {\bf A} \cdot d {\bf l}$, around the hexagonal unit cell equals the magnetic flux per unit cell, $\phi$, is legitimate, we take a particular choice of the gauge where only one of the three $\phi_{ij}$'s adjoining the nearest-neighbor carbon pairs is set to be non-zero.
The situation is illustrated in Fig.~\ref{fig:optimal_gauge}.
This gauge is called the optimal gauge since  the size of the magnetic unit cell (MUC) is optimal with its value being $q S_0$ for $\phi/\phi_0=p/q$, where $S_0$ is the area of a single hexagonal unit cell~\cite{Hatsugai1,Hasegawa2,Halperin}.
Note that the size of the magnetic unit cell is doubled in the usual Landau gauge~\cite{Zhang,Goerbig1}.

In the optimal guage, Harper's equations can be written as follows:
\begin{align}
E\psi_{\alpha n}^\A  &= \psi_{\alpha, n-1}^\B +\psi_{\alpha n}^\B +e^{2\pi i n \frac{\phi}{\phi_0}}\psi_{\alpha+1, n-1}^\B \;,\label{eq:HarperA1}  \\
E\psi_{\alpha n}^\B  &= \psi_{\alpha, n+1}^\A +\psi_{\alpha n}^\A +e^{-2\pi i (n+1) \frac{\phi}{\phi_0}} \psi_{\alpha-1, n+1}^\A \;,
\label{eq:HarperB1}
\end{align}
where $\alpha$ denotes the position of a given magnetic unit cell along the $y$-direction and the dimer index, $n$, indicates the position of each dimer within the magnetic unit cell.
The size of the magnetic unit cell is determined by the magnetic lattice translation symmetry.
For a rational value of the magnetic flux per unit cell in units of magnetic flux quantum, $\phi/\phi_0=p/q$,
Harper's equations in Eq.~(\ref{eq:HarperA1}) and (\ref{eq:HarperB1}) become periodic with respect to the diagonal lattice translation operation of $n \rightarrow n +l q$ with $l$ being an arbitrary integer.
Thus, in this situation, the magnetic unit cell covers the dimer index ranging from $n_0$ to $n_0+q-1$ with $n_0$ being an arbitrary initial dimer index. See Fig.~\ref{fig:optimal_gauge} for illustration.

Harper's equations in Eq.~(\ref{eq:HarperA1}) and (\ref{eq:HarperB1}) can be simplified by using the lattice translation symmetry along the $y$-direction.
That is to say, the $\alpha$-dependence can be removed by defining the crystal momentum, $k_y$, via the Bloch theorem, $\psi_{\alpha n} = \psi_{n \tilde{k}_y} e^{i \tilde{k}_y \alpha}$, with $\tilde{k}_y=k_y \sqrt{3}a$.
In this representation, Harper's equations are given by
\begin{align}
E\psi_{n \tilde{k}_y}^\A &= A_n(\tilde{k}_y)\psi^\B_{n-1, \tilde{k}_y} +\psi^\B_{n \tilde{k}_y}  ,                  \label{eq:HarperA2}\\
E\psi^\B_{n \tilde{k}_y} &= A_{n+1}^*(\tilde{k}_y)\psi^\A_{n+1, \tilde{k}_y} +\psi^\A_{n \tilde{k}_y} ,       \label{eq:HarperB2}
\end{align}
where
\begin{equation}
A_n(\tilde{k}_y)=2 e^{i\left(n\pi \frac{\phi}{\phi_0} +\frac{\tilde{k}_y}{2}\right)}\cos{\left(n\pi \frac{\phi}{\phi_0} +\frac{\tilde{k}_y}{2}\right)} .
\label{eq:A_n}
\end{equation}

\begin{figure}
\includegraphics[width=1\columnwidth]{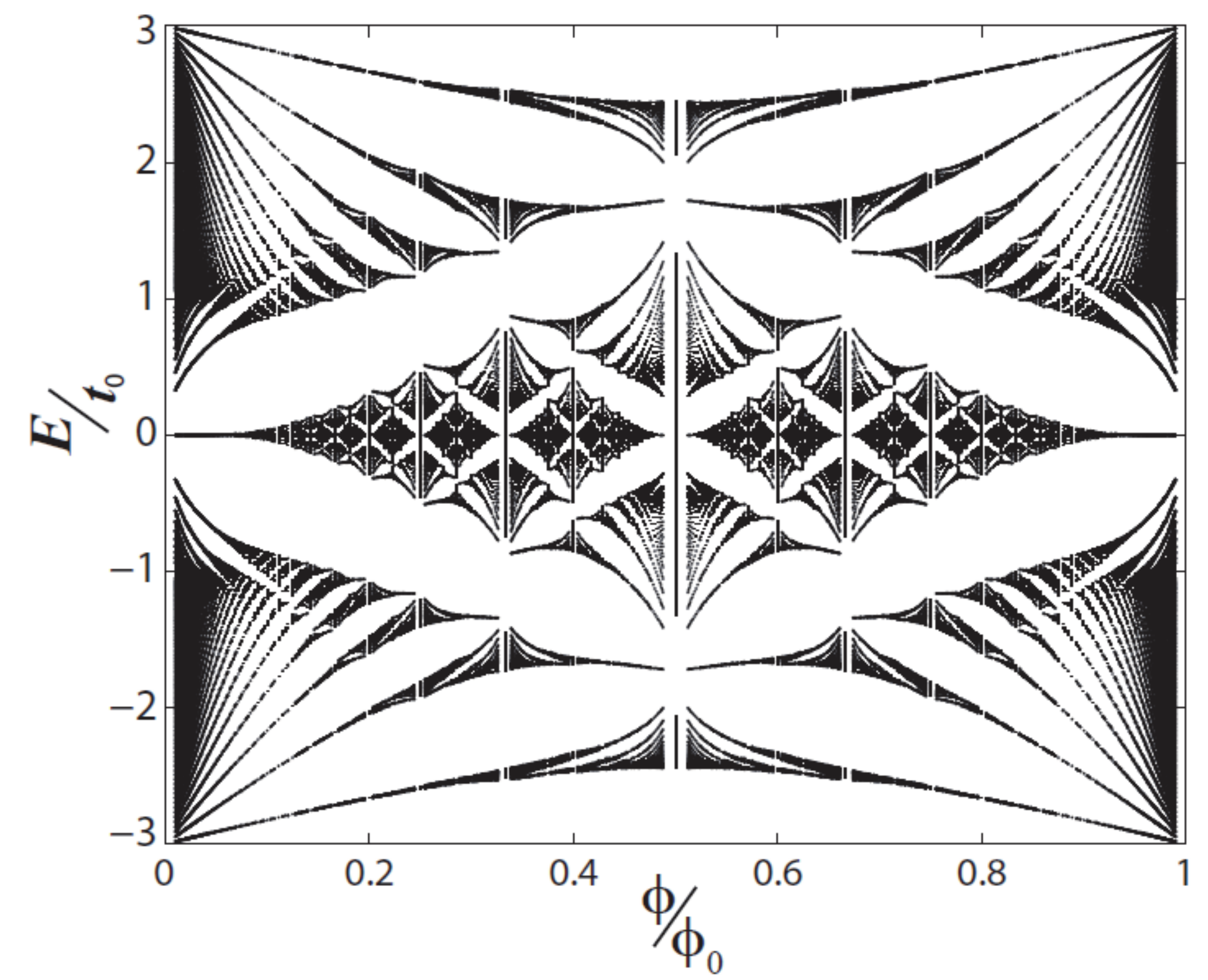}
\caption{(Color online) Hofstadter butterfly showing the energy eigenvalue, $E/t_0$, as a function of the magnetic flux per unit cell in units of magnetic flux quantum, $\phi/\phi_0$. Here, $t_0$ is the hopping amplitude in the absence of external magnetic field.}
\label{fig:full_Hofstadter_butterfly}
\end{figure}

By realizing that the Bloch condition along the diagonal direction, $\psi_{n \tilde{k}_y}= e^{i \tilde{k}_d n} \phi_{\tilde{k}_d \tilde{k}_y}(n)$ with $\phi_{\tilde{k}_d \tilde{k}_y}(n)$ being a periodic function of $n$ with period $q$, is equivalent to the boundary condition, $\psi_{n+q, \tilde{k}_y}= e^{i \tilde{k}_d q} \psi_{n \tilde{k}_y}$,
one can convert Harper's equations to an eigenvalue problem of the following $2q\times 2q$ Hamiltonian matrix:
\begin{widetext}
\begin{eqnarray}
\label{eq:matrix}
\mathbf{H}=
\begin{pmatrix}
 0                  & 1                &                      &                 &                     &                 &           &           &             & A_{n_0} e^{-i \tilde{k}_d q}\\
 1                  & 0                & A^*_{n_0+1} &                 &                     &                  &          &           &             &                 \\
                     & A_{n_0+1} & 0                   & 1              &                     &                 &           &           &             &                 \\
                     &                  & 1                   & 0              & A^*_{n_0+2} &                 &           &           &             &                 \\
                     &                  &                     & A_{n_0+2} & 0                 &                  &          &           &             &                 \\
                     &                  &                     &                  &                     &   \cdots     &          &           &             &                 \\
                     &                  &                     &                  &                    &                  & 0       & 1         &             &                 \\
                     &                  &                     &                  &                   &                   & 1       & 0         & A^*_{n_0+q-1} &                 \\
                     &                  &                     &                  &                   &                  &           & A_{n_0+q-1} &  0           & 1               \\
 A^*_{n_0}e^{i \tilde{k}_d q} &      &                    &                  &                   &                   &           &                      & 1            & 0
\end{pmatrix} ,
\label{eq:H}
\end{eqnarray}
\end{widetext}
where $n_0$, the first dimer index for a given magnetic unit cell, can be chosen arbitrarily since the choice of $n_0$ does not affect the energy eigenvalue.
Note that $\tilde{k}_d$ is the diagonal momentum measured in units of $1/\sqrt{3}a$.
Figure~\ref{fig:full_Hofstadter_butterfly} shows all energy eigenvalues of the above Hamiltonian matrix as a function of the magnetic flux per unit cell, $\phi$, in units of magnetic flux quantum, $\phi_0$.
This diagram is known as the Hofstadter butterfly.
Note that our result is completely identical to the previous result obtained by Rammal using the Landau gauge~\cite{Rammal}.

\section{Zero-energy mode}
\label{section:zero-energy}

While every energy eigenvalue of the Azbel-Hofstadter problem can be in principle found numerically, the size of the Hamiltonian matrix, that needs to be diagonalized, diverges as $2q \times 2q$ when $q \rightarrow \infty$ in the weak-field limit of $\phi/\phi_0=p/q$ with fixed $p$.
Obviously, a better approach is necessary in the weak-field regime.
In this work, we present an effective Hamiltonian method that can be used to provide an accurate analytic description of the central band of the Hofstadter butterfly in the weak-field regime.

Evolving from the $n=0$ Landau level, the central Hofstadter band (CHB) is most intriguing since it may hold a key to the mysteries of the fractional quantum Hall effect (FQHE) in graphene.
Note that, while the fractional quantum Hall effect has been observed in graphene, its detailed properties are not yet fully consistent with current theoretical understanding~\cite{Andrei,PKim2,PKim3,PKim4,Yacoby,Chakraborty,Mac1,Jain1,Jain2,Nomura1,Nomura2,Goerbig2,Jain3,Mac2}.
For one thing, the excitation energy gap, which is the most essential physical observable determining the electron transport, is orders-of-magnitude smaller than the corresponding theoretical predictions.
While this discrepancy could be explained by various perturbations such as disorder, Landau-level mixing, or ripples of the graphene layer, it is believed that the conclusive explanation for its true origin is still missing.
We think that a precise understanding of the nature of the central Hofstadter band can serve as an important step towards achieving such explanation.

Our effective Hamiltonian method is based on the observation that (i) all energy eigenstates of the central Hofstadter band are well approximated by those of the zero energy, which we call the zero-energy modes, and thus (ii) a very accurate effective Hamiltonian can be constructed by generating basis wave functions from the zero-energy modes.
In order to facilitate the discussion for how to construct the effective Hamiltonian, let us first investigate various properties of the zero-energy modes in this section.
Actual construction of the basis wave functions is performed in Sec.~\ref{section:effective_Hamiltonian}.

For $E=0$, Harper's equations in Eq.~(\ref{eq:HarperA2}) and (\ref{eq:HarperB2}) become decoupled between sublattice A and B:
\begin{align}
\psi_n^\A &= \psi_{n_0}^\A\prod_{m=n_0+1}^n \left[ -\frac{1}{A^*_m(\tilde{k}_y)} \right] ,          \label{eq:zero-modeA}\\
\psi_n^\B &= \psi_{n_0}^\B\prod_{m=n_0+1}^n \left[ -A_m(\tilde{k}_y) \right] ,                        \label{eq:zero-modeB}
\end{align}
where $\psi_{n_0}^\A$ and $\psi_{n_0}^\B$ (which are the amplitudes of the wave function at $n=n_0$ for sublattice A and B, respectively) can be regarded as simple normalization constants.
Seemingly otherwise, Eq.~(\ref{eq:zero-modeA}) and (\ref{eq:zero-modeB}) are not yet the solutions for Harper's equations since the momentum is not specified.
The momentum is fixed by imposing the boundary condition, $\psi_{n+q}= e^{i\tilde{k}_d q} \psi_{n}$ (which is due to the Bloch theorem).
The situation is a bit unorthodox here since the computation is performed in reverse order to the conventional scheme where the energy eigenvalue is determined for a given momenum.
In the current scheme, we seek for the right momentum corresponding to the zero-energy solution.

To find the right momentum for the zero-energy mode, it is convenient to use the following cosine product identity:
\begin{align}
\prod_{m=n+1}^{n+q} \cos{\left( m\pi\frac{p}{q} +\alpha \right)}  =\frac{e^{i\pi\gamma_{pqn}}}{2^{q-1}} \sin{ \left( \left(\alpha+\frac{\pi}{2}\right) q \right)},
\label{eq:cosine_product}
\end{align}
where $\gamma_{pqn}=pn+1+(q+1)(p-1)/2$.
The derivation of the cosine product identity is given in Appendix~\ref{appendix:cosine_product}.
By using the cosine product identity, one can simplify $\psi^\B_{n+q}/\psi^\B_{n}$ as follows:
\begin{align}
\frac{\psi^\B_{n+q}}{\psi^\B_{n}}
&= \prod^{n+q}_{m=n+1} \left[   -A_m (\tilde{k}_y)  \right]
\nonumber \\
&=  \prod^{n+q}_{m=n+1}  \left[ -2 e^{i\left(m\pi \frac{p}{q} +\frac{\tilde{k}_y}{2}\right)}\cos{\left(m\pi \frac{p}{q} +\frac{\tilde{k}_y}{2}\right)} \right]
\nonumber \\
&= 2e^{i\pi\delta_{pqn}} e^{i\frac{q\tilde{k}_y}{2}} \sin{ \left( (\tilde{k}_y +\pi) \frac{q}{2} \right)} ,
\label{eq:ratio1}
\end{align}
where $\delta_{pqn}=2np+(p+1)(q+1)-(q+1)/2$.
By noting that $2np$ and $(p+1)(q+1)$ are always even integers with $p$ and $q$ being coprime, one can re-write Eq.~(\ref{eq:ratio1}) as follows:
\begin{align}
\frac{\psi^\B_{n+q}}{\psi^\B_{n}}
&= 2e^{-i\frac{\pi}{2}(q+1)} e^{i\frac{q\tilde{k}_y}{2}} \sin{ \left( (\tilde{k}_y +\pi) \frac{q}{2} \right)} .
\label{eq:ratio2}
\end{align}

Then, the boundary condition, $\psi_{n+q}/\psi_{n} = e^{ik_d q}$, gives rise to the following equation for the zero-energy mode momentum:
\begin{align}
2\sin{\left(  (\tilde{k}_y+\pi)\frac{q}{2} \right)} = e^{i\tilde{k}_d q -i\frac{q\tilde{k}_y}{2} +i\frac{\pi}{2}(q+1)} ,
\label{eq:zero-mode1}
\end{align}
from which $\tilde{k}_y$ and $\tilde{k}_d$ can be simultaneously determined.
First, noting that the magnitude of the left-hand side should be unity, one can determine $\tilde{k}_y$ by imposing
\begin{align}
\sin{\left(  (\tilde{k}_y+\pi)\frac{q}{2} \right)}=\frac{(-1)^j}{2}
\label{eq:zero-mode2}
\end{align}
with $j$ being an integer.
The solution of Eq.~(\ref{eq:zero-mode2}), $\tilde{k}_y^*$, is given by:
\begin{align}
\tilde{k}_y^*=
\begin{cases}
\frac{\pi}{3q}-\pi+\frac{2\pi}{q}j \\
\\
\frac{5\pi}{3q}-\pi+\frac{2\pi}{q}j
\end{cases} .
\label{eq:k_y}
\end{align}
Then, by inserting Eq.~(\ref{eq:k_y}) into (\ref{eq:zero-mode1}), one can determine the other momentum for the zero-energy mode, $\tilde{k}_d^*$, whose value is given as follows:
\begin{align}
\tilde{k}_d^*=
\begin{cases}
\frac{5\pi}{3q}-\pi+\frac{2\pi}{q}l \\
\\
\frac{\pi}{3q}-\pi+\frac{2\pi}{q}l
\end{cases} ,
\label{eq:k_d}
\end{align}
with $l$ being an integer.
Note that, while the preceding computation is performed only for sublattice B, it can be shown that the zero-energy momentum is exactly the same for  sublattice A as well.
So far, the conclusion is that the wave function for the zero-energy mode is precisely described by Eq.~(\ref{eq:zero-modeA}) and (\ref{eq:zero-modeB}) with the appropriate momenta given by Eq.~(\ref{eq:k_y}) and (\ref{eq:k_d}).

\begin{figure*}
\includegraphics[width=1.8\columnwidth]{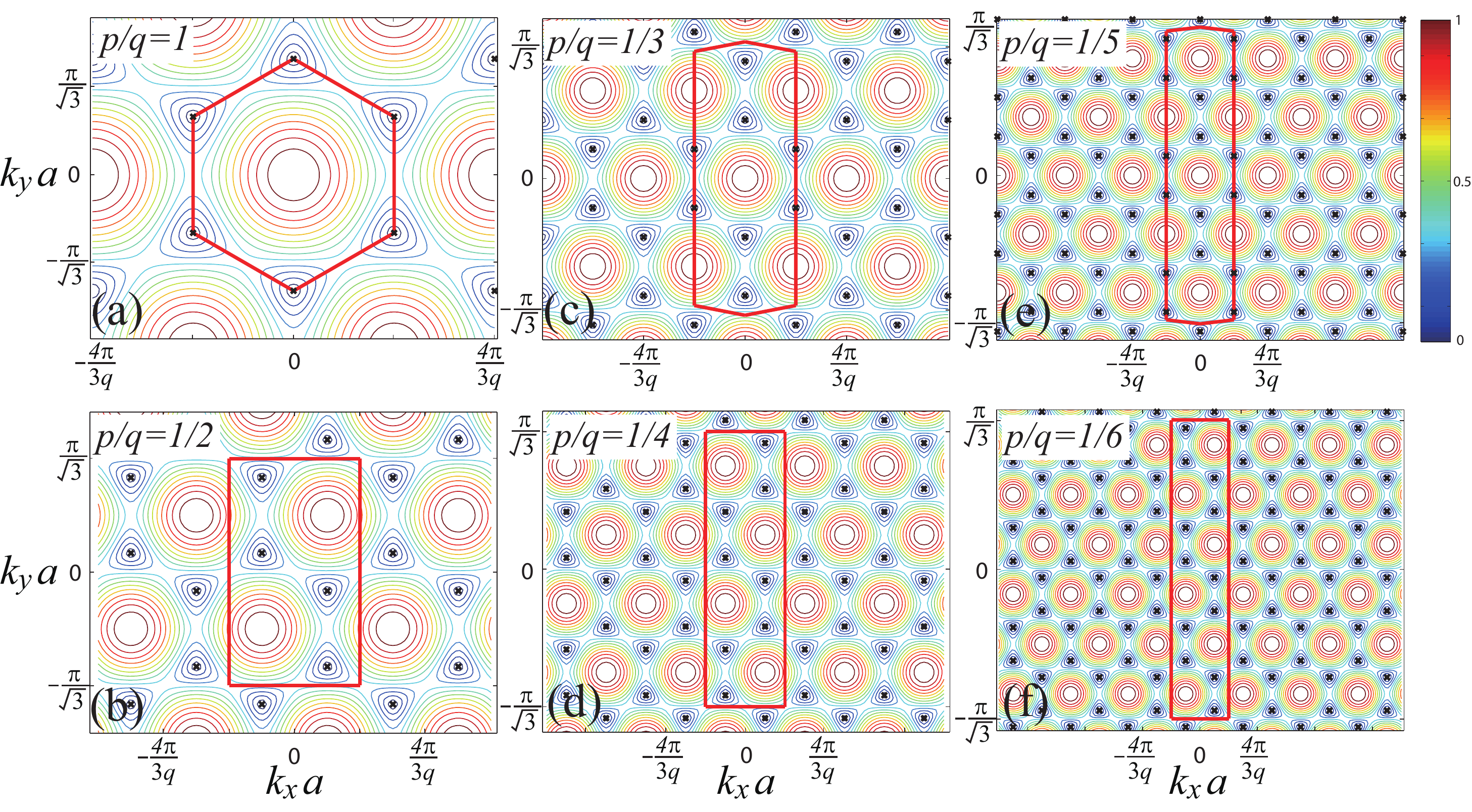}
\caption{(Color online) Contour plots for the energy dispersion at various flux values of $\phi/\phi_0=p/q$, with $p=1$ and $q$ increasing from 1 to 6 between panel (a) through (f).
In the figure, the energy dispersions are normalized by their respective half bandwidth.
The positions of the zero-energy momenta are denoted by little x marks and the magnetic Brillouin zones are enclosed by red solid lines.
The above energy dispersions are computed by solving either the original Harper's equations or the effective Hamiltonian method explained in Sec.~\ref{section:effective_Hamiltonian}, both of which produce essentially the identical results.
It is interesting to note that the effective Hamiltonian method works well even for $p/q=1$ owing to the mirror structure of the Hofstadter butterfly, which maps the region near $p/q=1$ to the weak-field counterpart.}
\label{fig:Dirac_cones}
\end{figure*}

At this point, it is illuminating to obtain the locations of the zero-energy momenta in the magnetic Brillouin zone.
To this end, let us convert $\tilde{k}_d$ in terms of the usual cartesian coordinates.
Since $\tilde{k}_d$ is the diagonal momentum along which the dimer index, $n$, increases within a given magnetic unit cell index, $\alpha$, the conversion rule is given by
\begin{align}
\tilde{k}_y &= \sqrt{3}a k_y , \nonumber \\
\tilde{k}_d &= \sqrt{3}a k_d = \frac{3}{2} a k_x +\frac{\sqrt{3}}{2} a k_y ,
\label{eq:conversion_rule}
\end{align}
which, combined with Eq.~(\ref{eq:k_y}) and (\ref{eq:k_d}), gives rise to to the following:
\begin{align}
k_y^* &=
\begin{cases}
\frac{1}{\sqrt{3}a} \left(  \frac{\pi}{3q} -\pi +\frac{2\pi}{q} j \right)  \\
\\
\frac{1}{\sqrt{3}a} \left(  \frac{5\pi}{3q} -\pi +\frac{2\pi}{q} j \right)
\end{cases} ,
\nonumber \\
k_x^* &= \frac{1}{a} \left(  \frac{\pi}{q} -\pi +\frac{2 \pi}{3q} l'       \right) ,
\end{align}
where $l'=2l-j$.
Figure~\ref{fig:Dirac_cones} presents the energy dispersions of the central Hofstadter band in the form of contour plot for various flux values, where the positions of the zero-energy momenta are denoted by little x marks.
As one can see, the zero-energy momenta occur exactly in the same honeycomb pattern as the Dirac points in the absence of magnetic field.
Actually, it is shown in Sec.~\ref{subsection:continuum_limit} and \ref{subsection:general_flux} that, in the weak-field limit, with proper energy and momentum re-scaling, the energy dispersion of the central Hofstadter band becomes exactly identical to that of graphene in the absence of magnetic field, proving that the zero-energy modes are, in fact, nothing but massless Dirac particles.
The energy dispersion remains very close to that in the absence of magnetic field even when the magnetic flux per unit cell becomes moderately large.

It is interesting to mention that the number of zero-energy modes is given by $2q$ within each magnetic Brillouin zone and this fact is related with the Landau-level degeneracy of graphene in the continuum limit.
In the lattice model, the degeneracy of the Landau level can be regarded as the number of different ways of locating the wave packet maximum within the magnetic unit cell.
Since the magnetic unit cell contains $2q$ carbon atoms, the wave packet maximum can have $2q$ different locations and therefore the so-defined Landau-level degeneracy is $2q$, which, in the continuum limit, becomes infinite, or a macroscopic number proportional to the system size.

\begin{figure}
\includegraphics[width=0.8\columnwidth]{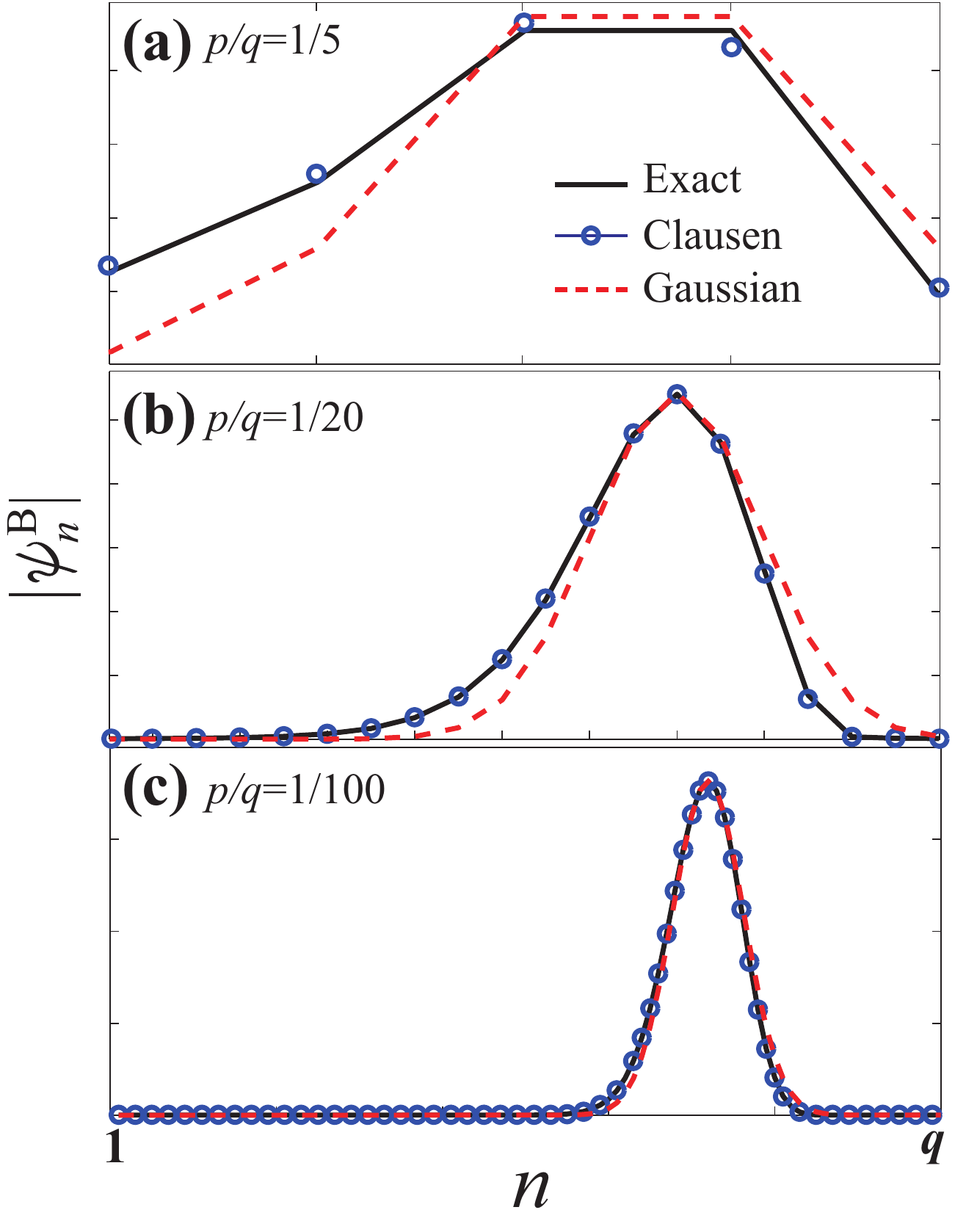}
\caption{(Color online) Wave function profiles for the zero-energy mode as a function of dimer index, $n$, at three different flux values: (a) $\phi/\phi_0=p/q=1/5$, (b) $1/20$, and (c) $1/100$.
As one can see, at moderate flux values, say, $p/q=1/5$ and $1/20$, there is a sizable asymmetry around the maximum position.
The asymmetry is seen more clearly in contrast to the Gaussian wave packet (red dashed lines) which is the exact energy eigenstate in the continuum, or weak-field, limit.
It is important to note that, while the Gaussian wave packet provides a poor representation of the exact results (solid lines) at moderate flux values, a new analytic expansion method using the Clausen function (open circles) works very well for a wide range of flux values.}
\label{fig:wavefunc_profile}
\end{figure}

We now investigate the wave function profile for the zero-energy mode.
The wave function for the zero-energy mode can be computed numerically by using Eq.~(\ref{eq:zero-modeA}) and (\ref{eq:zero-modeB}).
Figure~\ref{fig:wavefunc_profile} shows the results for several different flux values.
One of the most salient features of the exact wave function profile is the fact that it is asymmetric around its maximum point while, in the continuum limit, the zero-energy wave function reduces to the Gaussian wave packet (which is the energy eigenstate in the $n=0$ Landau level) and therefore should be symmetric.
As one can see from Fig.~\ref{fig:wavefunc_profile}, however, the deviation from the Gaussian shape vanishes rather rapidly as the flux per unit cell decreases.

Actually, in the weak-field regime, it is possible to derive a better analytic approximation for the zero-energy wave function than the simple Gaussian.
The basic idea is, first, to convert the zero-energy wave function represented in a product form to a summation form by taking the logarithm and, then, to approximate the summation with an integral by regarding, $x_n \equiv n\pi\phi/\phi_0+\tilde{k}_y/2$, as a continuous variable.
This procedure is valid when $\phi/\phi_0$ is small.
Relegating the detailed derivation to Appendix~\ref{appendix:Clausen_derivation}, here, we present the final result:
\begin{align}
\left|\psi^\A_{n}\right| &\propto \exp{\left[ \frac{1}{2\pi\phi/\phi_0} \mathrm{Cl}_2 \left( 2\pi\frac{\phi}{\phi_0} n+\eta \right) \right]}  , \nonumber \\
\left|\psi^\B_{n}\right| &\propto \exp{\left[ -\frac{1}{2\pi\phi/\phi_0} \mathrm{Cl}_2 \left( 2\pi\frac{\phi}{\phi_0} n+\eta \right) \right]} ,
\label{eq:Clausen_wave_packet}
\end{align}
where $\eta=\tilde{k}_y+\pi(\phi/\phi_0+1)$ and $\textrm{Cl}_2 (\theta)$, called the Clausen function, is defined such that  $\textrm{Cl}_2(\theta)=\sum^{\infty}_{n=1} \sin{(n\theta)}/n^2$.
From now on, let us call the wave function profile given by Eq.~(\ref{eq:Clausen_wave_packet}) the Clausen wave packet.
As one can see from Fig.~\ref{fig:wavefunc_profile}, the Clausen wave packet provides a very accurate approximation of the exact results for a wide range of flux values.

To confirm analytically that the Clausen wave packet indeed reduces to the Gaussian in the continuum limit, it is convenient to use the Landau gauge, in which case the Clausen approximation corresponds to the following:
\begin{align}
\left|\psi^\B_{n+1}\right| \propto \exp{\left[ -\frac{1}{2\pi\phi/\phi_0} \mathrm{Cl}_2 \left( 2\pi\frac{\phi}{\phi_0} n+\kappa \right) \right]} ,
\label{eq:Clausen_Landau}
\end{align}
where $\kappa=\tilde{k}_y -\frac{5\pi}{3}\phi/\phi_0+\pi$.
Here, we only consider the amplitudes in sublattice B since the same process can be applied to sublattice A.
With the definition of a new continuous variable, $x=\frac{3a}{2}(n-1)$ (where $a$ is the lattice constant), Eq.~(\ref{eq:Clausen_Landau}) can be re-written as follows:
\begin{align}
\left|\psi^{\B}(x)\right| \propto \exp{\left[-\frac{l^2_\B}{S_0}\mathrm{Cl}_2\left(\frac{\sqrt{3}a}{l^2_\B}x   +\kappa'  \right)\right]} ,
\label{eq:Clausen_of_x}
\end{align}
where $2\pi\phi/\phi_0=S_0/l^2_\B$, with $S_0=\frac{3\sqrt{3}}{2}a^2$ being the area of the hexagonal unit cell, is used.
In the above, $\kappa'=\kappa+S_0/l^2_\B$.

Now, noting that Eq.~(\ref{eq:Clausen_of_x}) is maximized when the Clausen function inside the exponential becomes minimized, we search for the condition minimizing $\mathrm{Cl}_2(\sqrt{3}ax/l^2_\B+\kappa')$.
To this end, it is convenient to use the following property of the Clausen function that $\mathrm{Cl}_2(\theta)$ has a maximum at $\theta=\pi/3$ and a minimum at $5\pi/3$ within a single period between 0 and $2\pi$.
Note that  $\mathrm{Cl}_2(\theta)$ is a periodic function with $2\pi$ period.
Then, one can determine the maximum position of $|\psi^\B(x)|$ as follows:
\begin{align}
x_\textrm{max}(\mu)=-q_yl^2_\B-\frac{1}{4}a+\frac{l^2_\B}{\sqrt{3}a/2}\mu\pi
\end{align}
where $\mu$ is an integer and $q_y=k_y+\frac{4\pi}{3\sqrt{3}a}$ is the difference between $k_y$ and the momentum of one of the two Dirac points.
(Note that, for sublattice A, $q_y$ is defined as the difference between $k_y$ and the momentum of the other Dirac point.)
Since the Clausen function can be expanded around its minimum positions as follows:
\begin{align}
\mathrm{Cl}_2(\theta) = \mathrm{Cl}_2(5\pi/3) +\frac{\sqrt{3}}{4} (\theta-\theta_\textrm{min})^2 +\cdots,
\label{eq:Clausen_expansion}
\end{align}
where $\mathrm{Cl}_2(5\pi/3)=-1.0149$ and $\theta_\textrm{min}=5\pi/3+2\mu\pi$ with $\mu$ being an integer,
the next step is to expand the Clausen wave packet in the vicinity of $x_\textrm{max}$, assuming $|x-x_\textrm{max}|/a \ll  l^2_\B/S_0$. 
It is important to note that such expansion becomes very accurate when the inverse coefficient in front of the Clausen function, $l^2_\B/S_0$, is much larger than the deviation of the Clausen function from its minimum position. 
The expansion is given as follows:
\begin{align}
-\frac{l^2_\B}{S_0} \mathrm{Cl}_2 \left(\frac{\sqrt{3}a}{l^2_\B}x +\kappa' \right) \approx
-\frac{l^2_\B}{S_0} \left[\lambda_0+\lambda_2 (x-x_\textrm{max})^2 \right] ,
\end{align}
where the linear term vanishes due to the extremum condition.
As shown from the comparison with Eq.~(\ref{eq:Clausen_expansion}),
the zeroth-order coefficient, $\lambda_0$, is equal to $\mathrm{Cl}_2(5\pi/3)$ and
the second-order coefficient is given by
$\lambda_2 = \frac{\sqrt{3}}{4} (\sqrt{3}a/l^2_\B)^2=S_0/(2l^4_\B)$, which finally gives rise to the desired result that the Clausen wave packet reduces to the usual Gaussian function of $\exp{(-(x-x_\textrm{max})^2/2l^2_\B)}$.
Note that this result is exactly the same as the previous result obtained by Goerbig and collaborators~\cite{Goerbig1}.

For later use, it is convenient to compute the maximum as well as the minimum positions of the zero-energy wave function for the optimal gauge in the weak-field regime.
In the case of sublattice B, the maximum (minimum) position arises whenever the cosine factor of $A_n(\tilde{k}_y)$ in Eq.~(\ref{eq:A_n}), $\left|\cos{(n\pi\frac{\phi}{\phi_0}+\frac{\tilde{k}_y}{2})}\right|$,  passes through $1/2$ from above (below) to below (above) as a function of dimer index, $n$.
Note that $n\pi \frac{\phi}{\phi_0}$ can be treated roughly as a continuous variable so long as $\phi/\phi_0$ is sufficiently small.
With the maximum and the minimum position denoted as $n^\B_\textrm{max}$ and $n^\B_\textrm{min}$, respectively, the result is as follows:
\begin{align}
n^\B_{\textrm{max},s} &=\mathrm{floor}\left[\frac{1}{\pi\phi/\phi_0} \left(\frac{\pi}{3}-\frac{\tilde{k}_y}{2}+s\pi\right) \right]  ,     \nonumber \\
n^\B_{\textrm{min},s}  &=\mathrm{floor}\left[\frac{1}{\pi\phi/\phi_0} \left(\frac{2\pi}{3}-\frac{\tilde{k}_y}{2}+s\pi\right) \right] ,    \label{eq:maxmin}
\end{align}
where $s$ is an integer.
In the case of sublattice A, it can be shown that $n^\A_{\textrm{max}}=n^\B_{\textrm{min}}$ and $n^\A_{\textrm{min}}=n^\B_{\textrm{max}}$ since the cosine factor is multiplied inversely in this case.
Finally, it is interesting to mention that, in the strong-field regime where the magnetic flux is in the vicinity of unity, i.~e., $|\phi/\phi_0 -1| \ll 1$, the maximum and the minimum-position formula is modified as follows:
\begin{align}
n^\B_{\textrm{max},s} &=\mathrm{floor}\left[\frac{1}{\pi(1-\phi/\phi_0)} \left(\frac{\pi}{3}+\frac{\tilde{k}_y}{2}+s\pi\right) \right]  ,     \nonumber \\
n^\B_{\textrm{min},s}  &=\mathrm{floor}\left[\frac{1}{\pi(1-\phi/\phi_0)} \left(\frac{2\pi}{3}+\frac{\tilde{k}_y}{2}+s\pi\right) \right] ,    \label{eq:maxmin2}
\end{align}
where $s$ is, again, an integer. 

\section{Effective Hamiltonian}
\label{section:effective_Hamiltonian}

In the preceding section, we have carefully investigated various aspects of the zero-energy solution for Harper's equations.
Despite many nice, analytic properties, the zero-energy modes alone consist of only a negligible part of the entire magnetic Brillouin zone.
While all energy eigenvalues can be, in principle, computed by solving Harper's equations, a brute-force numerical diagonalization is prohibited in the weak-field regime where the size of the Hamiltonian matrix quickly diverges.
To scan the entire Brillouin zone in the weak-field region, it is necessary to devise a better method.
In this section, we present such a method using the effective Hamiltonian, which provides a very accurate description of the central Hofstadter band in the entire Brillouin zone.

\subsection{Basis wave functions}
\label{subsection:basis_wavefunc}

The essence of our effective Hamiltonian method lies in choosing the right set of basis wave functions most relevant to the central Hofstadter band.
To do so, it is important to note that, for $\phi/\phi_0=p/q$, the central Hofstadter band always contains $2p$ subbands.
One way of understanding this is, first, to realize that the wave function profile of all energy eigenstates comprising the central Hofstadter band is more or less identical to that of the zero-energy modes in the weak-field limit.
Then, from Eq.~(\ref{eq:maxmin}), one can see that there should be exactly $p$ local maxima for the wave function profile inside the magnetic unit cell (while their individual maximum values can be different).
Now, imagine that $\tilde{k}_y$ increases from 0 to $2\pi$ so that the entire Brillouin zone is covered along the $y$-direction.
According to Eq.~(\ref{eq:maxmin}), this process is actually identical to decreasing $s$ by unity, which in turn means that that the wave function is translated exactly by one unit of the distance between the nearest maxima.
This process covers only $1/p$ of the whole magnetic unit cell.
To fill the whole magnetic unit cell, $p$ bands are necessary.
Since the same is true for both sublattice A and B, there should be $2p$ subbands for the central Hofstadter band.

\begin{figure*}
\includegraphics[width=1.9\columnwidth]{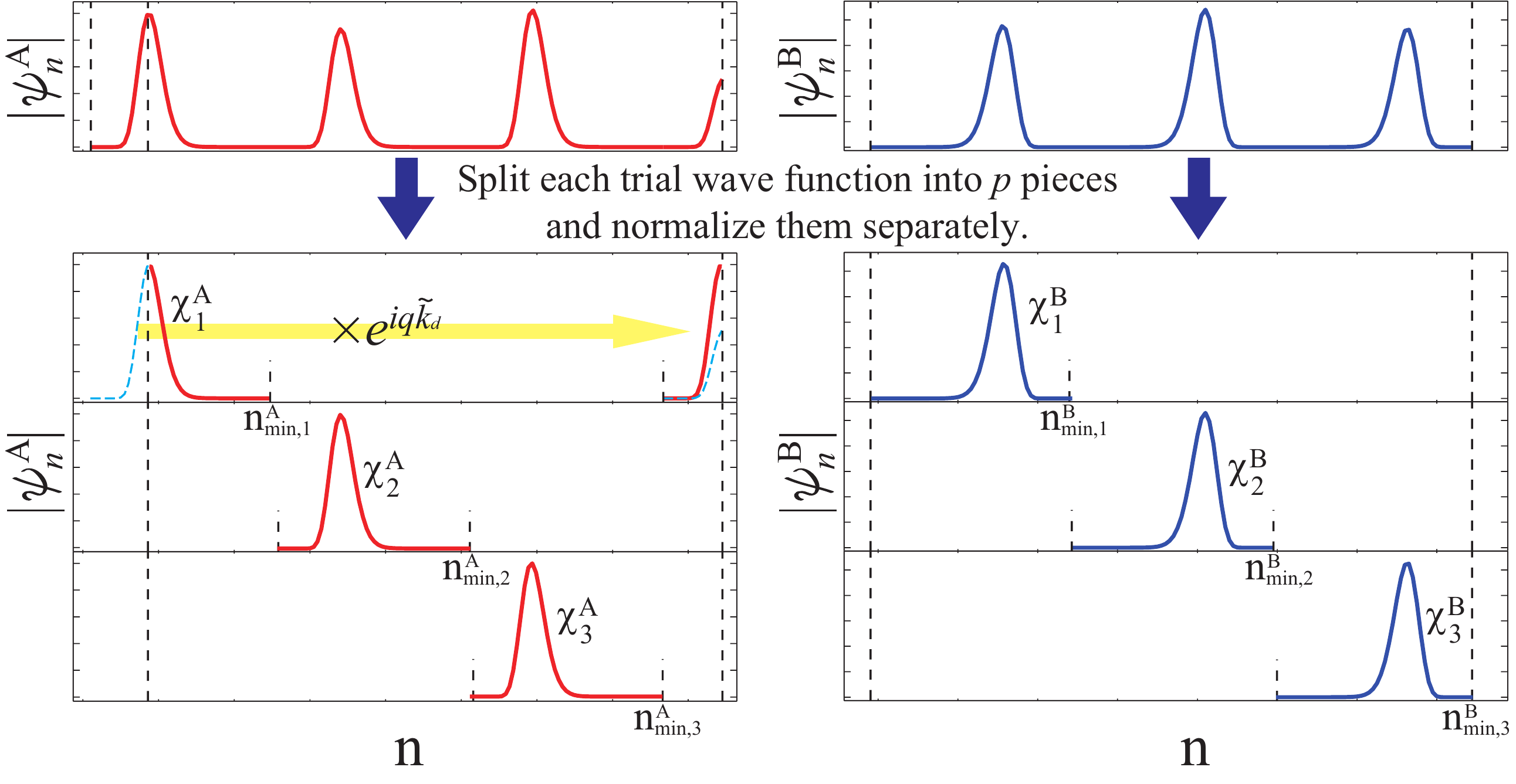}
\caption{(Color online) Schematic diagram for the construction of basis wave functions in the case of $p/q=3/152$.
Basis wave functions for the central Hofstadter band can be constructed in three steps:
(i) for a given momentum, ${\bf k}=(k_x,k_y)$, one generates a trial basis wave function according to the zero-energy formula in Eq.~(\ref{eq:zero-modeA}) and (\ref{eq:zero-modeB}), (ii) then, slices the so-obtained trial wave function into equally-spaced $p$ pieces such that each piece contains exactly one local maximum, and (iii) finally, normalizes the $p$ piece-wise basis wave functions separately for each sublatttice A and B.
Refer the text for details.}
\label{fig:basis_construction}
\end{figure*}

Now, we present a scheme for systemically constructing approximate, but very accurate basis wave functions for such $2p$ subbands.
This basis-constructing scheme is best explained in the following three steps.
(i) First, for a given momentum, ${\bf k}=(k_x,k_y)$, we compute a trial basis wave function by using the zero-energy formula in Eq.~(\ref{eq:zero-modeA}) and (\ref{eq:zero-modeB}).
For the time being, let us ignore normalization.
(ii) We then slice the so-obtained trial wave function into equally-spaced $p$ pieces such that each piece contains exactly one local maximum in the region located between two consecutive minima of the trial basis wave function.
Care must be taken for sublattice A where, according to our convention, the boundary of the magnetic unit cell sits right on top of one of the wave function maxima and thus the piece-wise basis wave function containing such maximum is split into two regions separated across the magnetic unit cell.
In this case, to satisfy the periodic boundary condition dictated by the Bloch theorem, we multiply an additional phase factor, $e^{i\tilde{k}_d q}$, to the copied portion of the wave function amplitude translated from the outside to the ending part of the magnetic unit cell.
(iii) By normalizing the $p$ piece-wise basis wave functions separately for each sublattice, we finally obtain $2p$ basis wave functions.
Note that the finally obtained basis wave functions are orthonormal to each other.
See Fig.~\ref{fig:basis_construction} for an illustration of the basis-constructing scheme.

Explicitly, the basis wave function for sublattice B, $\chi^\B_{s}(n)$ with $s$ ranging from 1 to $p$, can be written as follows:
\begin{widetext}
\begin{align}
\chi^\B_{s}(n)=
\begin{cases}
c^\B_s \prod_{m=n^\B_{\textrm{min},s-1}+1}^n  \left[-A_{m}(\tilde{k}_y)\right]    &  \textrm{for $n^\B_{\textrm{min}, s-1} < n \leq n^\B_{\textrm{min}, s}$},  \\
\\
0    & \mathrm{otherwise},
\end{cases}
\label{eq:basisB}
\end{align}
 \end{widetext}
where $c^\B_s$ is the normalization constant.
Note that $\chi^\B_{s}(n)$ is the piece-wise basis wave function containing the $s$-th maximum.
For sublattice A, the situation is similar except for the special case of $s=1$ where the wave function maximum is split into two regions across the magnetic unit cell:
\begin{widetext}
\begin{align}
\chi^\A_{1}(n)=
\begin{cases}
c^\A_1 \prod_{m=n_0+1}^n  \left[-1/A^*_{m}(\tilde{k}_y)\right]    &  \textrm{for $n_0 < n \leq n^\A_{\textrm{min}, 1}$},  \\
\\
e^{i\tilde{k}_d q} c^\A_1 \prod_{m=n^\A_{\textrm{min},p-1}+1}^n  \left[-1/A^*_{m}(\tilde{k}_y)\right]    &  \textrm{for $n^\A_{\textrm{min},p} < n \leq n_0+q-1$},  \\
\\
0    & \mathrm{otherwise},
\end{cases}
\label{eq:basisA1}
\end{align}
\end{widetext}
where $n_0$ is the first dimer index in the magnetic unit cell, which, according to our convention, is $n^\B_{\textrm{min},0}+1$.
Note that the last dimer index is $n_0 +q-1$, which is in turn equal to $n^\B_{\textrm{min},p}$.
In the above, $c^\A_1$ is the normalization constant.
For the other cases with $s \neq 1$, the formula is given similarly to that of sublattice B:
\begin{widetext}
\begin{align}
\chi^\A_{s}(n)=
\begin{cases}
c^\A_s \prod_{m=n^\A_{\textrm{min},s-1}+1}^n  \left[-1/A^*_{m}(\tilde{k}_y)\right]    &  \textrm{for $n^\A_{\textrm{min}, s-1} < n \leq n^\A_{\textrm{min}, s}$},
\\
0    & \mathrm{otherwise},
\end{cases}
\label{eq:basisAs}
\end{align}
 \end{widetext}
where, again, $c^\A_s$ is the normalization constant.

\subsection{Constructing the effective Hamiltonian}
\label{subsection:constructing_H_eff}

The basic idea behind our effective Hamiltonian method is to isolate the Hilbert space near zero energy in terms of the basis wave functions constructed in the preceding section.
With $p$ number of basis wave functions for each sublattice A and B, say, $\chi^\A_\mu$ and $\chi^\B_\nu$ with $\mu, \nu= 1,\cdots, p$, our Hamiltonian can be written as a $2p \times 2p$ matrix as follows:
\begin{align}
\mathbf{H}^{\mathrm{eff}} =
\begin{pmatrix}
\mathbf{0}                             &                     \mathbf{H}^{\A\B}                   \\
{\mathbf{H}^{\A\B}}^\dag      &                     \mathbf{0}
\end{pmatrix} ,
\label{eq:H_eff}
\end{align}
where $\mathbf{H}^{\A\B}$ is a $p\times p$ matrix whose elements are given by
\begin{align}
(\mathbf{H}^{\A\B})_{\mu\nu} = \left< \chi^\A_\mu \right|  \mathbf{H} \left| \chi^\B_\nu   \right> .
\label{eq:matrix_element_of_H_eff}
\end{align}
In the above, $\mathbf{H}$ is the original Hamiltonian matrix for Harper's equations given in Eq.~(\ref{eq:H}).
Note that all elements in the block-diagonal part of $\mathbf{H}^\mathrm{eff}$ are strictly zero since $\mathbf{H}$ allows only the nearest-neighbor hopping.

\subsection{Approaching the continuum limit along $\phi/\phi_0=1/q$}
\label{subsection:continuum_limit}

The effective Hamiltonian takes the most compact form in the case of $\phi/\phi_0=1/q$.
The reason is that, in this case, there is only a single basis wave function for each sublattice and thus the size of the effective Hamiltonian becomes just $2\times 2$ no matter how large $q$ may become.
In fact, it is important to note that  the larger $q$ becomes, the more accurate results our effective Hamiltonian method provides, as shown later in this section.
In addition to the mathematical simplicity, the case of $\phi/\phi_0=1/q$ is physically important since taking the large-$q$ limit along $\phi/\phi_0=1/q$ is one of the most natural paths approaching the continuum limit, via which the central Hofstadter band evolves into the $n=0$ Landau level.

With all diagonal elements vanishing (for the reason explained in the preceding section), the only non-zero, off-diagonal elements of the $2\times 2$ effective Hamiltonian are $(\mathbf{H}^{\A\B})_{11}$ and its complex conjugate:
\begin{align}
(\mathbf{H}^{\A\B})_{11} &= \left< \chi^\A_1 \right|  \mathbf{H} \left| \chi^\B_1   \right>
\nonumber \\
&=\left(\chi^{\A}_{1,n_0} \right)^*  \left( \chi^{\B}_{1,n_0}+A_{n_0}e^{-i\tilde{k}_d q} \chi^\B_{1,n_0+q-1} \right)  \nonumber\\
&={\cal C} \left\{ 1-e^{-i\tilde{k}_d q} \prod_{m=n_0}^{n_0+q-1} \left[ -A_m(\tilde{k}_y) \right] \right\}
\nonumber \\
&={\cal C}  \left\{ 1+e^{-i\tilde{k}_d q} \left[ e^{i\tilde{k}_d q} -e^{-i \pi q}  \right]   \right\}
\label{eq:off-diag}
\end{align}
where ${\cal C}=(\chi^{\A}_{1,n_0})^*\chi^{\B}_{1,n_0}$ and the cosine product identity in Eq.~(\ref{eq:ratio2}) is used to obtain the last line.
The step connecting between the first and the second line of Eq.~(\ref{eq:off-diag}) indicates that only a single term from the inner product survives.
This is due to the fact that all the other terms vanish strictly by the very definition of the basis wave functions given in Eq.~(\ref{eq:basisB}), (\ref{eq:basisA1}), and (\ref{eq:basisAs}), which, in the case of $\phi/\phi_0=1/q$, is simply identical to the zero-energy formula in Eq.~(\ref{eq:zero-modeA}) and (\ref{eq:zero-modeB}) due to the fact that there is only a single maximum in the magnetic unit cell in this case.

Diagonalizing the $2\times 2$ effective Hamiltonian gives rise to the following energy eigenvalues,
\begin{widetext}
\begin{align}
E_{1/q}^\pm(\vec{k}) = \pm |{\cal C}| \sqrt{1+4\cos^2{\left( (\tilde{k}_y-\tilde{k}_d)\frac{q}{2}\right)} -4(-1)^q
\cos{\left( (\tilde{k}_y-\tilde{k}_d)  \frac{q}{2}\right)}
\cos{\left( (\tilde{k}_y+\tilde{k}_d)  \frac{q}{2}\right)}  } ,
\label{eq:dispersion}
\end{align}
\end{widetext}
as well as the corresponding eigenstates,
\begin{align}
\Phi_{1/q}^\pm=\frac{1}{\sqrt{2}}\left( \chi^\A_1 \pm e^{-i\theta} \chi^\B_1 \right) ,
\label{eq:eigenstate}
\end{align}
where $\theta$ is defined such that $e^{i\theta}=(\mathbf{H}^{\A\B})_{11}/|(\mathbf{H}^{\A\B})_{11}|$
It is interesting to note that the energy eigenstates are always composed of an equal mixture between sublattice A and B.

\begin{figure}
\includegraphics[width=0.9\columnwidth]{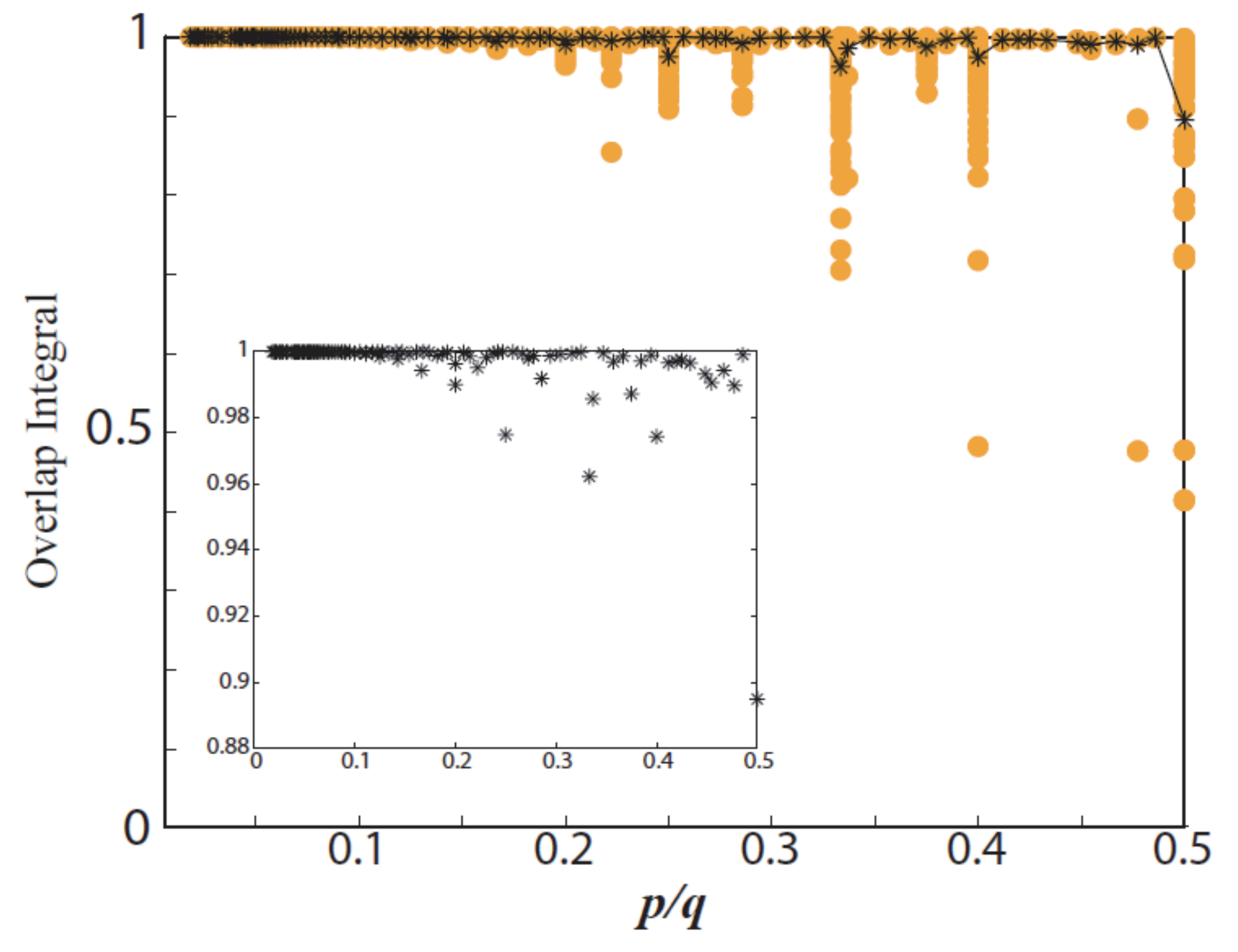}
\caption{(Color online) Overlap integral between the energy eigenstates obtained from the effective Hamiltonian method and the exact counterparts from the original Harper's equations.
Stars indicate the averaged value of the overlap integral over all crystal momenta within the magnetic Brillouin zone, while circles denote individual results for different momenta.
It is important to note that the overlap integral approaches unity very rapidly as $\phi/\phi_0=p/q$ decreases.}
\label{fig:overlap}
\end{figure}

Figure~\ref{fig:overlap} shows evidence for the validity of the effective Hamiltonian method in terms of the overlap integral between the eigenstates obtained from the effective Hamiltonian and the exact counterparts from the original Harper's equations.
As one can see, the overlap is very close to unity for all momenta at small flux values up to $p/q=0.2$. 
Actually, the overlap is not too bad all the way up to $p/q=0.5$ when averaged over all crystal momenta within the magnetic Brillouin zone. 
Note that, for general flux values of $p/q$, the effective energy eigenstates are obtained by solving the $2p \times 2p$ effective Hamiltonian.
See Sec.~\ref{subsection:general_flux} for details.

To get more physical insight on the energy dispersion in Eq.~(\ref{eq:dispersion}), it is convenient to convert $\tilde{k}_d$ in terms of the usual cartesian coordinates as done previously in Eq.~(\ref{eq:conversion_rule}).
The result is quite illuminating:
\begin{widetext}
\begin{align}
E_{1/q}^\pm(\vec{k})
&=\begin{cases} \pm |{\cal C}| \sqrt{1+4\cos^2{\left(\frac{\sqrt{3}}{2}qa k_y\right)}+4\cos{\left(\frac{\sqrt{3}}{2}qa k_y\right)}
\cos{\left(\frac{3}{2}qa k_x \right) } } & (q:\mathrm{odd})\\ \\
\pm |{\cal C}| \sqrt{1+4\cos^2{\left(\frac{\sqrt{3}}{2}q (a k_y-\frac{\pi}{\sqrt{3}q} )\right)}
+4\cos{\left(\frac{\sqrt{3}}{2}q (a k_y-\frac{\pi}{\sqrt{3}q} )\right)}
\cos{\left(\frac{\sqrt{3}}{2}q (a k_x-\frac{\pi}{3q} )\right)} } & (q:\mathrm{even})
\end{cases} ,
\label{eq:dispersion_cartesian}
\end{align}
\end{widetext}
which shows that, with proper energy and momentum re-scaling, the energy dispersion is, in fact, exactly identical to that in the absence of magnetic field.
Note that, for $q$ even, the momentum is shifted by $\Delta {\bf k}=(\frac{\pi}{3qa},\frac{\pi}{\sqrt{3}qa})$.
The above energy dispersions were plotted in the form of contour graph previously in Fig.~\ref{fig:Dirac_cones} for various flux values, which shows explicitly that massless Dirac particles exist in the central Hofstadter band.

\begin{figure}
\includegraphics[width=0.8\columnwidth]{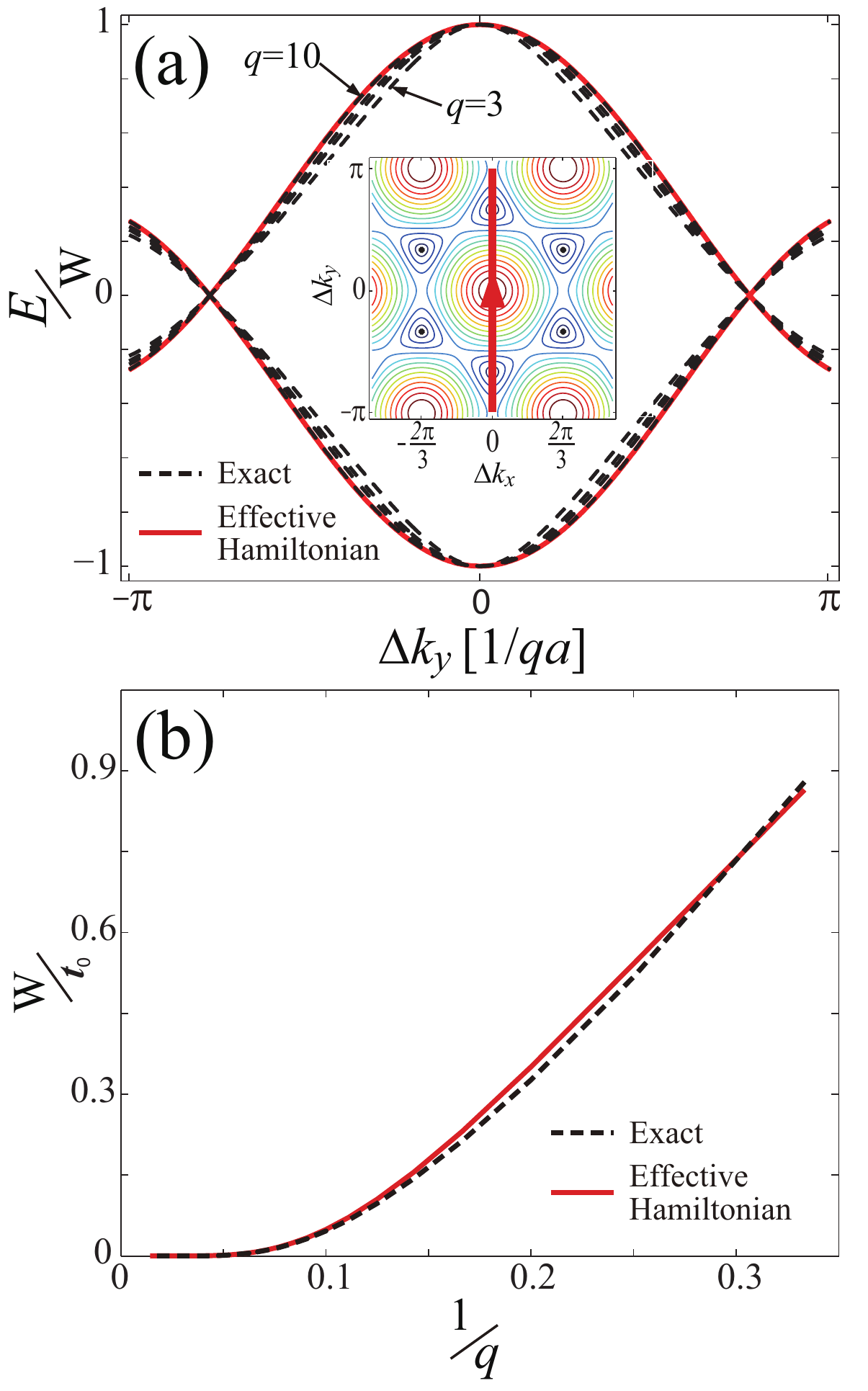}
\caption{(Color online) (a) Comparison between the exact energy dispersions obtained from the original Harper's equations (black dashed lines) and that from the effective Hamiltonian method (red solid line) for various $\phi/\phi_0=1/q$.
Note that, with proper energy and momentum rescaling, all energy dispersions obtained from the effective Hamiltonian at different $\phi/\phi_0=1/q$ collapse into a single curve.
In the figure, the energy dispersions are normalized by their respective half bandwidth, $W$, and the momentum is expressed in units of $1/qa$.
The inset shows the path in the magnetic Brillouin zone, along which the momentum is scanned.
Note that the scanning path is chosen such that it passes through the Dirac points. 
(b) Comparison between the exact half bandwidth and that obtained from the effective Hamiltonian method as a function of $\phi/\phi_0=1/q$.}
\label{fig:1_q_flux}
\end{figure}

Figure~\ref{fig:1_q_flux} shows a detailed comparison between the exact energy dispersions obtained from the original Harper's equations and that from the effective Hamiltonian method for various $\phi/\phi_0=1/q$.
It is important to note that, with proper energy and momentum re-scaling, all energy dispersions obtained from the effective Hamiltonian collapse into a single curve.
In the figure, the momentum is expressed in units of $1/qa$ and the energy dispersion is normalized by the half bandwidth, $W$, which is related with the prefactor, ${\cal C}$, via $W=3|{\cal C}|$.
As one can see from Fig.~\ref{fig:1_q_flux}~(a), the agreement between the exact results for the normalized energy dispersion and that from the effective Hamiltonian method is quite good for $q$ as small as 3 and becomes perfect quickly as $q$ increases.
In addition to the re-scaled shape of the energy dispersion, it is shown below that the bandwidth of the energy dispersion itself is also captured extremely accurately by the effective Hamiltonian method.

To determine the bandwidth of the energy dispersion, it is necessary to compute the prefactor, ${\cal C}$, in Eq.~(\ref{eq:dispersion}):
\begin{align}
|{\cal C}| =  |\chi^\A_{1,n_0}| |\chi^\B_{1,n_0}|  \approx |{\cal C}_\A| |{\cal C}_\B| ,
\end{align}
where the Clausen approximation for the zero-energy wave function in Eq.~(\ref{eq:Clausen_wave_packet}) is used:
\begin{align}
|\chi^\A_{1,n_0}| &\approx |{\cal C}_\A| e^{\frac{1}{2\pi\phi/\phi_0} \mathrm{Cl}_2 \left( 2\pi\frac{\phi}{\phi_0} n_0+\eta \right)} ,
\nonumber \\
|\chi^\B_{1,n_0}| &\approx |{\cal C}_\B| e^{-\frac{1}{2\pi\phi/\phi_0} \mathrm{Cl}_2 \left( 2\pi\frac{\phi}{\phi_0} n_0+\eta \right)} .
\end{align}
Here, ${\cal C_\A}$ and ${\cal C_\B}$ are the normalization constants for sublattice A and B, respectively.
Note that $\eta=\tilde{k}_y+\pi(\phi/\phi_0+1)$ can be regarded as just a constant for the current purposes.

We now need to compute the normalization constants, ${\cal C_\A}$ and ${\cal C_\B}$.
First, due to the sublattice symmetry, $|{\cal C_\A}|=|{\cal C_\B}|$, and therefore $|{\cal C}|=|{\cal C_\B}|^2$ .
Mathematically, this is a consequence of the property of the Clausen function:  $-\mathrm{Cl}_2(\theta)=\mathrm{Cl}_2 (2\pi-\theta)$.
Second, with the substitution of $\theta= 2\pi\frac{\phi}{\phi_0} n +\eta$, the normalization condition can be approximated by the following integral form:
\begin{align}
1 &= |{\cal C_\B}|^2 \sum_{n=n_0}^{n_0+q-1} e^{-\frac{1}{\pi\phi/\phi_0} \mathrm{Cl}_2 \left( 2\pi\frac{\phi}{\phi_0} n+\eta \right)}
\nonumber \\
&\approx |{\cal C_\B}|^2 \frac{1}{2\pi\phi/\phi_0} \int^{2\pi}_0 d\theta e^{-\frac{1}{\pi\phi/\phi_0}  \mathrm{Cl}_2 (\theta)}
\nonumber \\
&\approx |{\cal C_\B}|^2 \frac{1}{2\pi\phi/\phi_0} \int^{\infty}_{-\infty} d\theta e^{-\frac{1}{\pi\phi/\phi_0}  \mathrm{Cl}_2 (5\pi/3)-\frac{l^2_\B}{3a^2} \theta^2} ,
\end{align}
where the last line is obtained in the limit of small $\phi/\phi_0$, in which the integrand becomes sharply peaked around the minimum position of the Clausen function occurring at $\theta=5\pi/3$ [see Eq.~(\ref{eq:Clausen_expansion})].
In this limit, it is also safe to extend the integral range to $(-\infty,\infty)$.
Following is the final result for the half width of the central Hofstadter band, $W/t_0$:
\begin{align}
\frac{W}{t_0}= 3|{\cal C}| \approx 3^{5/4} \sqrt{\phi/\phi_0} \exp{\left[\frac{1}{\pi\phi/\phi_0}\mathrm{Cl}_2(5\pi/3)\right]} ,
\label{eq:band_width}
\end{align}
where we have re-introduced the hopping amplitude, $t_0$, for convenience.
Figure~\ref{fig:1_q_flux}~(b) shows the comparison between the exact half bandwidth and that from the effective Hamiltonian method in Eq.~(\ref{eq:band_width}) as a function of $\phi/\phi_0=1/q$, which, as one can see, are in excellent agreement.
It is interesting to note that, in units of the energy level spacing between the $n=1$ and 0 Landau level, $\Delta=\sqrt{2} \hbar v_F/l_\B$, the half width of the central Hofstadter band becomes simplified as follows:
\begin{align}
\frac{W}{\Delta} = \frac{3}{\sqrt{2\pi}}  \exp{\left[\frac{1}{\pi\phi/\phi_0}\mathrm{Cl}_2(5\pi/3)\right]} ,
\end{align}
where $\mathrm{Cl}_2(5\pi/3)=-1.0149$.

\subsection{General flux}
\label{subsection:general_flux}

\begin{figure}
\includegraphics[width=1\columnwidth]{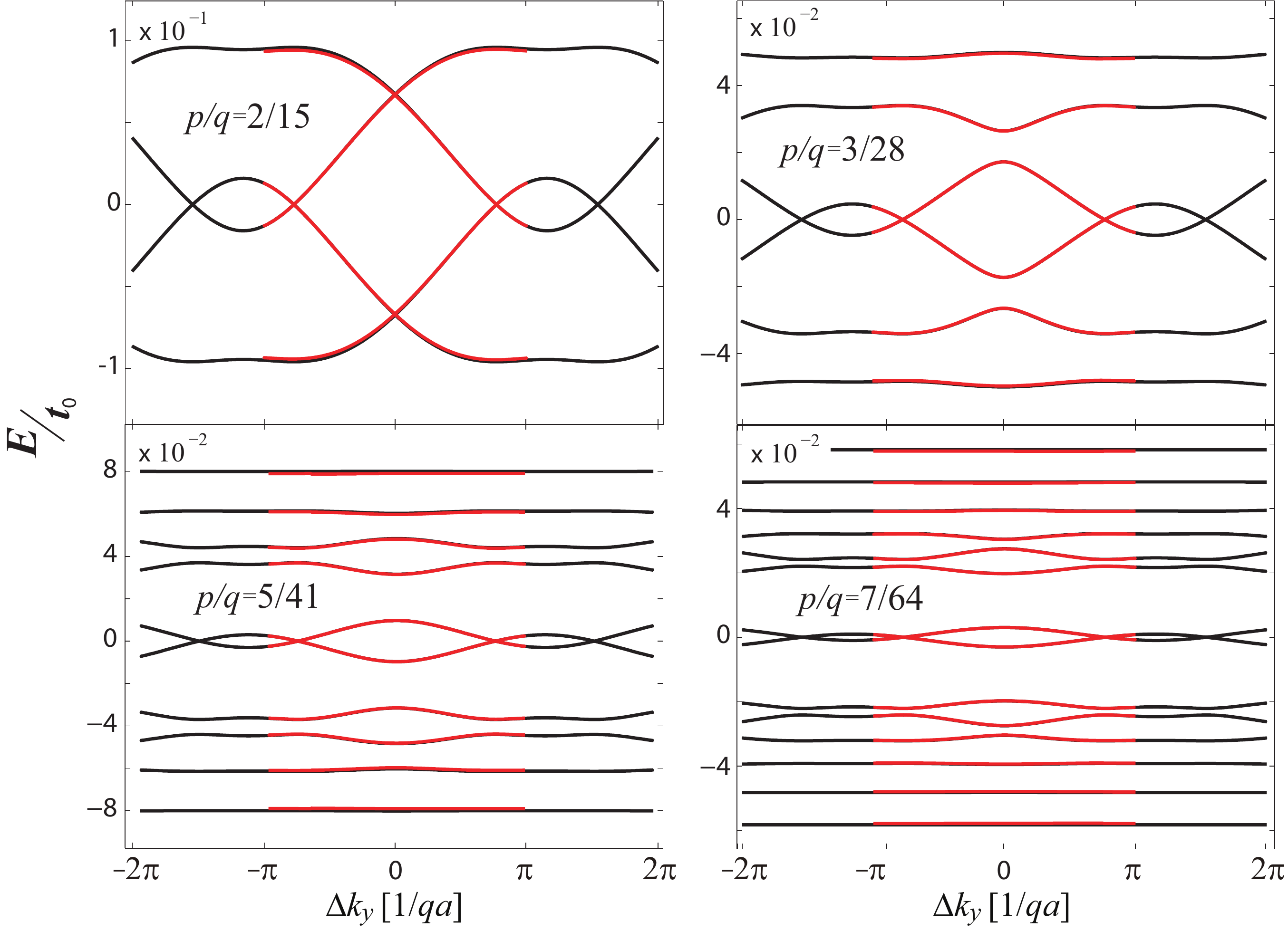}
\caption{(Color online) Comparison between the energy dispersions obtained from the original Harper's equations (black lines) and those from the effective Hamiltonian method (red lines) at various $\phi/\phi_0=p/q$.
For clarity, the energy dispersions from the effective Hamiltonian method are plotted only within the window of $-\pi/qa \leq \Delta k_y \leq \pi/qa$.
Note that the momentum is scanned along the same path as in Fig.~\ref{fig:1_q_flux}~(a).
\label{fig:general_flux}}
\end{figure}

At general flux, $\phi/\phi_0=p/q$, the mathematical expression for the energy eigenvalue as well as eigenstate are not as simple as those at $\phi/\phi_0=1/q$, which are given by Eq.~(\ref{eq:dispersion}) and (\ref{eq:eigenstate}), respectively, in the preceding section.
Nevertheless, it is emphasized that, for $p/q \ll 1$, the size of the effective Hamiltonian, which is $2p \times 2p$, is much reduced from that of the original Harper's equation, which is $2q \times 2q$.
This means that the fine self-similar structures of the central Hofstadter band in the weak-field regime can be computed in a much efficient manner. 
As shown in the following section, this, combined with some analytic results obtained at $\phi/\phi_0=1/q$, in turn enables us to make a prediction that massless Dirac particles should occur under arbitrary magnetic field.

Postponing the detailed discussion to the following section, here, we present the comparison between the results obtained from the effective Hamiltonian method and those from the original Harper's equations for general $\phi/\phi_0=p/q$. 
Figure~\ref{fig:general_flux} provides numerical results for the energy dispersion at various flux values in comparison with those from the effective Hamiltonian method.
As one can see, the agreement is excellent not only for the bands near zero energy, but also for the entire $2p$ bands within the central Hofstadter band.

\section{Self-similar occurrence of massless Dirac particles}
\label{section:self-similar}

\begin{figure}
\includegraphics[width=0.6\columnwidth]{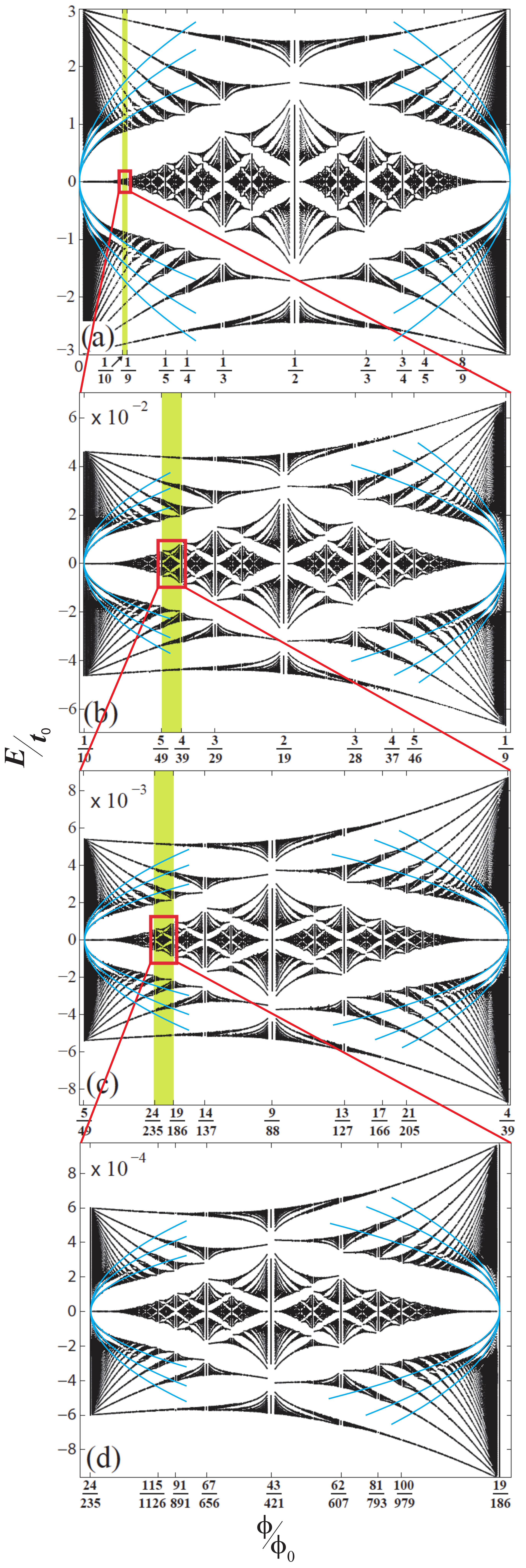}
\caption{(Color online) A sequence of zoomed views for the Hofstadter butterfly in graphene showing various self-similar recursive patterns.
Note that a fan of narrow energy bands are emanated from each single-band boundary flux (SBF), $\phi_\textrm{SBF}$, which,
as indicated by blue guiding curves, scale as $\textrm{sgn}(n)\sqrt{|n(\phi-\phi_\textrm{SBF})|}$ with $n$ being an integer. This scaling behavior is a signature of the formation of recursive Landau levels associated with self-similarly occurring massless Dirac particles.
}
\label{fig:Hofstadter_zoom}
\end{figure}

It is mentioned in the preceding section that the effective Hamiltonian method can help reveal the fine self-similar structures of the central Hofstadter band in the weak-field regime much efficiently.
The results obtained from the effective Hamiltonian method are shown in Fig.~\ref{fig:Hofstadter_zoom}, which provides a sequence of zoomed views unveiling the self-similar recursive patterns of the central Hofstadter band.

One of the most salient features of the Hofstadter butterfly seen in Fig.~\ref{fig:Hofstadter_zoom} is that the central Hofstadter band is partitioned by a series of special flux values, at which the central Hofstadter band is entirely composed of a single band appearing as a vertical line in the figure.
Note that, for example, in the top-most panel of Fig.~\ref{fig:Hofstadter_zoom}, the central Hofstadter band is partitioned by a series of vertical lines occurring at $\phi/\phi_0=1/q$ and $1-1/q$ with $q$ being a positive integer.
Similar patterns are observed in subsequently zoomed views.
For later convenience, we name the flux belonging to this series of special flux values as the single-band boundary flux (SBF).

In fact, owing to the self-similar recursive pattern of the central Hofstadter band, it is convenient to coin the name the ``$n$-th level'' central Hofstadter band (CHB) and the corresponding ``$n$-th level'' single-band boundary flux (SBF).
In this nomenclature, the first level SBF values are given by $1/q$ and $1-1/q$ with $q$ being a positive integer.
Meanwhile, the second panel of Fig.~\ref{fig:Hofstadter_zoom} shows that the second level SBF values are given by $5/49$, $4/39$, $3/29$, $2/19$, $3/28$, $4/37$, $5/46$, and so on.
The third level SBF values can be determined similarly from the third panel.
At this point, it is natural to ask the question if there is a rule for the SBF values and, if so, what mathematical form it takes.

The answer is that, indeed, there is a single rule for all SBF values, the mathematical form of which is given as follows:
\begin{equation}
f  =s_1 +\cfrac{(-1)^{s_1}}{n_1 +s_2
+\cfrac{(-1)^{s_2}}{n_2 +s_3
+\cfrac{(-1)^{s_3}}{n_3 +\cdots}
}} \;\;\;,
\label{eq:SBF}
\end{equation}
where $n_i$, a positive integer larger than 2, and $s_i$, either 0 or 1, are determined by the following recursion rule.
Suppose that $f$ is one of the SBF values.
Then, we first define $f_0=f$.
If $\mathrm{floor}{\left(1/f_0\right)} \geq 2$, we set $n_1=\mathrm{floor}{\left(1/f_0\right)}$ and $s_1=0$.
Otherwise, i.~e., if $\mathrm{floor}{\left(1/f_0\right)} = 1$, we set $n_1=\mathrm{floor}{\left[1/(1-f_0)\right]}$ and $s_1=1$.
As the next recursion step, we then define $f_1=1/f_0-n_1$ for the former and $1/(1-f_0)-n_1$ for the latter case.
We now repeat the same procedure to determine $n_2$ and $s_2$ from $f_1$.
This procedure can be continued until we get $f_{n}=0$ with $n$ indicating that $f$ is the $n$-th level SBF.

It is instructive to explain the above rule by using an example.
As an example, let us take $\phi/\phi_0=4/39$, which is one of the second-level SBF values.
According to the above-mentioned rule, we first define $f_0=4/39$.
Since $1/f_0=39/4=9+3/4$, $n_1=9$, $s_1=0$, and subsequently $f_1=3/4$.
Now that $1/f_1=4/3=1+1/3$, we have to set $n_2 =\mathrm{floor}[1/(1-f_1)]=4$, in which case $s_2=1$.
The recursion steps terminate at the second level since $f_2=0$.
In conclusion, $f=4/39$ can be expressed as follows:
\begin{align}
f= 4/39  = \cfrac{1}{9 +1
+\cfrac{(-1)}{4} }\;\;\;.
\end{align}
It is now convenient to devise a simplified notation scheme where the SBF is represented by a sequence of ${n_i}$ along with whether $s_i$ is 0 or 1.
One way of denoting the fact that $s_i=1$ is to put a bar on top of the corresponding $n_i$.
In this notation, $f=4/39=(9,\bar{4})$.
Similar computations can be performed to show that $f=19/186$ and $17/166$, which are among the third-level SBF values shown in the third panel in Fig.~\ref{fig:Hofstadter_zoom}, are represented by $(9, \bar{4}, \bar{4})$ and $(9, \bar{4}, 4)$, respectively.
On the other hand, $f=91/891$, which is one of the fourth-level SBF values shown in the fourth panel in Fig.~\ref{fig:Hofstadter_zoom}, is given by $(9,\bar{4},\bar{4},\bar{4})$.

By knowing the continued-fraction representation of a given SBF value, $f$,
one can extract two important pieces of information.
First, how many $n_i$'s exist indicates the level of $f$ as a SBF value.
Second, more importantly, provided that $f$ is the $m$-th level SBF,
$f$ is related to the first-level SBF occurring at $1/n_m$ (or $1-1/n_m$ via the reflection symmetry).
For example, $f=19/186=(9, \bar{4}, \bar{4})$ has four $n_i$'s and the last integer is 4, which tells us that $f=19/186$ is the fourth-level SBF related to the first-level SBF occurring $1/4$.

Once the relationship between a given SBF and its first-level counterpart is established, there is a far-reaching consequence.
To understand this, it is important to note that (i) the first-level SBF values are always either $1/q$ or $1-1/q$ with $q$ being a positive integer and (ii) for $\phi/\phi_0=1/q$ and $1-1/q$, the energy dispersion is isomorphic to that in the absence of magnetic field, as proven in Sec.~\ref{subsection:continuum_limit}.
Therefore, if all SBF values are related to their respective first-level counterparts, the energy dispersion at all SBF values should also be isomorphic to that in the absence of magnetic field.
In other words, massless Dirac particles should exist at all SBF values.
In fact, since all rational fractions can be represented by a continued fraction via Eq.~(\ref{eq:SBF}), massless Dirac particles should exist at all rational flux values.
This conclusion is supported by explicit numerical results obtained from both the original Harper's equations and the effective Hamiltonian method, which show that the energy dispersion is indeed isomorphic to that of graphene in the absence of magnetic field. 
This is, also, fully consistent with an analytic result that zero-energy modes always exist for general $\phi/\phi_0=p/q$ as shown in Sec.~\ref{section:zero-energy}. 
Moreover, since any irrational number can be represented as a continued fraction with an infinite number of levels, the energy dispersion at irrational flux values can be regarded as that of massless Dirac particles in the limit where the energy scale goes to zero. 
In this sense, we arrive at the final conclusion that, however small their energy scale may be, massless Dirac particles should exist at all flux values, rational or irrational.

A corollary of the above conclusion is that the central Hofstadter band should also contain a self-similar structure of recursive Landau levels associated with those self-similarly occurring massless Dirac particles. 
Figure~\ref{fig:Hofstadter_zoom} shows that each single-band boundary flux (SBF), $\phi_\textrm{SBF}$, indeed emanates a fan of narrow energy bands which, as indicated by blue guiding curves in the figure, scale as $\textrm{sgn}(n)\sqrt{|n(\phi-\phi_\textrm{SBF})|}$ with $n$ being an integer. This scaling behavior is a signature of the formation of recursive Landau levels.

\section{Conclusion}
\label{section:conclusion}

In this paper, we develop an effective Hamiltonian method that can be used to provide an accurate analytic description of the central Hofstadter band in graphene much more efficiently than directly solving the original Harper's equations in the weak-field regime.
The source of the efficiency is due to the fact that, in the weak-field regime where the magnetic flux per unit cell in units of magnetic flux quantum, $\phi/\phi_0=p/q \ll 1$, the size of the effective Hamiltonian is given by $2p \times 2p$, which is greatly reduced from that of the original Hamiltonian, $2q \times 2q$.
The benefit of using the effective Hamiltonian method is maximized at $\phi/\phi_0=1/q$, where the size of the effective Hamiltonian remains to be $2\times 2$ no matter how large $q$ may become.
Actually, the advantage of using the effective Hamiltonian is not simply due to the reduction of the matrix size, but rather the separation of the low-energy sector. 
It is important to note that solving the original Harper's equations generates unreliable, noisy data below certain small flux values where the low-energy sector becomes so narrow that the energy resolution falls below numerical accuracy.

By using such effective Hamiltonian method, we show explicitly that the energy dispersion is isomorphic to that in the absence of magnetic field for all flux values satisfying $\phi/\phi_0=1/q$, which in turn indicates that massless Dirac particles should exist no matter how small the magnetic flux may become.
In fact, by combing numerical results showing the self-similar recursive structure of the central Hofstadter band, we conclude that massless Dirac particles should occur under arbitrary magnetic flux.
If so, as a corollary, the central Hofstadter band should also contain a self-similar structure of recursive Landau levels.

As a useful by-product of the effective Hamiltonian method, we are also able to compute the width of the central Hofstadter band as a function of magnetic field, which can be used to assess the experimental feasibility of actually observing massless Dirac particles inside the central Hofstadter band.
In units of the energy level spacing between the $n=1$ and 0 Landau level, $\Delta=\sqrt{2}\hbar v_F / l_\B$, where $v_F$ is the Fermi velocity at Dirac point and $l_\B$ is the magnetic length, we show that the width of the central Hofstadter band is given by $W/\Delta=\frac{1}{\sqrt{2\pi}} \exp{(-\gamma \frac{\phi_0}{\phi})}$ with $\gamma= |\mathrm{Cl}_2(5\pi/3)|/\pi \simeq 0.323$.

Finally, we mention that the above effective Hamiltonian method is not applicable in the square lattice.
The reason is as follows.
The validity of the effective Hamiltonian method depends crucially on the fact that the zero-energy wave function has a well localized shape with exponentially negligible tails so that it can be safely split into linearly independent pieces with each forming the basis wave functions for the effective Hamiltonian.
No such simplification is possible in the square lattice where the zero-energy wave functions are extended all over the magnetic unit cell.
The situation is not improved in the case of non-zero energy states, whose wave function forms are no longer given by a simple product form and thus prohibit a systematic construction of the analytic basis wave functions from the outset.

\begin{acknowledgments}
This research was supported in part by the National Research Foundation of Korea (NRF) funded by the Korea government (MEST) under Quantum Metamaterials Research Center, Grant No. 2008-0062238 (K.P.).
Also, the authors thank KIAS Center for Advanced Computation for providing computing resources.
\end{acknowledgments}

\appendix

\section{Derivation of the cosine product identity}
\label{appendix:cosine_product}

In this section of Appendix, we prove the following cosine product identity:
\begin{align}
\prod_{m=n+1}^{n+q} \cos{\left( m\pi\frac{p}{q} +\alpha \right)}  =\frac{e^{i\pi\gamma_{pqn}}}{2^{q-1}} \sin{ \left( \left(\alpha+\frac{\pi}{2}\right) q \right)},
\label{appen_eq:cosine_product}
\end{align}
where $\gamma_{pqn}=pn+1+(q+1)(p-1)/2$.
Here, $p$ and $q$ are coprime natural numbers.

We begin by multiplying the both sides of Eq.~(\ref{appen_eq:cosine_product}) with $2^{q}$, in which case the left-hand side becomes
\begin{align}
&\prod_{m=n+1}^{n+q} \left[ 2 \cos{ \left(m\pi \frac{p}{q} +\alpha\right) } \right]                                                                                                   \nonumber \\
&= \prod_{m=n+1}^{n+q} \left[ e^{i \left(m\pi \frac{p}{q} +\alpha\right)} + e^{-i \left( m\pi \frac{p}{q} +\alpha\right)} \right]                                         \nonumber\\
&= \prod_{m=n+1}^{n+q} e^{-i \left(m\pi\frac{p}{q}  +\alpha\right)} \prod_{m=n+1}^{n+q} \left[ 1 + e^{i\left(2m\pi\frac{p}{q} + 2\alpha\right)} \right]  \nonumber\\
&= e^{-i\frac{\pi}{2} p(2n+q+1)-i\alpha q} \prod_{m=n+1}^{n+q} \left[ 1 + e^{i\left(2m\pi\frac{p}{q} +2\alpha\right)} \right] .
\label{appen_eq:left-hand}
\end{align}

Now, let us consider the product in the last line of Eq.~(\ref{appen_eq:left-hand}), $S= \prod_{m=n+1}^{n+q} \left[ 1 + e^{i\left(2m\pi\frac{p}{q} + 2\alpha\right)} \right]$, whose logarithm is written as follows:
\begin{align}
\ln{S} &= \sum_{m=n+1}^{n+q} \ln{  \left[ 1+ e^{i\left (2m\pi\frac{p}{q} +2\alpha \right)} \right]  }                     \nonumber\\
&= \sum_{m=n+1}^{n+q} \sum_{s=1}^{\infty} \frac{(-1)^{s-1}}{s} e^{i\left(2m\pi\frac{p}{q} +2\alpha\right)s} ,
\label{appen_eq:lnS1}
\end{align}
where the Taylor expansion of the logarithm, $\ln{(1+x)} = \sum_{s=1}^{\infty} \frac{(-1)^{s-1}}{s} x^s$, is used.
Note that the above Taylor expansion of the logarithm is valid for $|x| \leq 1$ with exception of $x=-1$.
This condition is satisfied for $x=e^{i\left(2m\pi\frac{p}{q} +2\alpha\right)}$ unless $e^{i\left(2m\pi\frac{p}{q} +2\alpha\right)}=-1$.
Fortunately, in the case when there is such $m$ that $e^{i\left(2m\pi\frac{p}{q} +2\alpha\right)}=-1$, Eq.~(\ref{appen_eq:cosine_product}) is automatically satisfied with the both sides becoming simultaneously zero.
The reason is that $e^{i\left(2m\pi\frac{p}{q} +2\alpha\right)}=-1$ means $2m\pi\frac{p}{q}+2\alpha=(2l-1)\pi$ with $l$ being an integer, which is in turn equivalent to
\begin{align}
\left(\alpha+\frac{\pi}{2}\right)q=(ql-pm)\pi.
\label{appen_eq:condition}
\end{align}
Since $ql-pm$ is an integer, the right-hand side of Eq.~(\ref{appen_eq:cosine_product}) becomes zero.
It is shown in Eq.~(\ref{appen_eq:left-hand}) that the left-hand side also vanishes when $e^{i\left(2m\pi\frac{p}{q} +2\alpha\right)}=-1$ .
Therefore, it can be concluded that the Taylor expansion in the above can be safely used.

Then, Eq.~(\ref{appen_eq:lnS1}) can be simplified as follows:
\begin{align}
\ln S &= \sum_{s=1}^{\infty} \frac{(-1)^{s-1}}{s} e^{i2\alpha s}
\sum_{m=n+1}^{n+q} e^{i 2m\pi \frac{p}{q} s}\nonumber\\
&= \sum_{s\neq 0~(\mathrm{mod}~q)} \frac{(-1)^{s-1}}{s} e^{i2\alpha s} \frac{1-e^{i 2\pi p s }}{1-e^{i 2\pi \frac{p}{q} s}} e^{i 2\pi \frac{p}{q} s(n+1)} \nonumber\\
&+
\sum_{s=0~(\mathrm{mod}~q)} \frac{(-1)^{s-1}}{s} e^{i2\alpha s}q \nonumber\\
&= \sum_{l} \frac{(-1)^{ql-1}}{ql} e^{i2\alpha ql}q \nonumber\\
&= \sum_{l} \frac{(-1)^{ql-1}}{l} e^{i2\alpha ql} \nonumber\\
&= \sum_{l} \frac{(-1)^{l-1}}{l} e^{i2\alpha ql} e^{i\pi(q-1)l} \nonumber\\
&= \sum_{l} \frac{(-1)^{l-1}}{l} \left\{ e^{i2 \left[ \alpha q +\frac{\pi}{2}(q+1) \right] }\right\}^l \nonumber\\
&= \ln \left\{1+ e^{i2 \left[ \alpha q +\frac{\pi}{2}(q+1) \right]} \right\} ,
\label{appen_eq:lnS2}
\end{align}
where the last step in the above equation is obtained when $e^{i2 \left[\alpha q +\frac{\pi}{2}(q+1) \right]} \neq -1$.
Fortunately, this condition is identical to the previous one that there is no such $m$ satisfying $e^{i\left(2m\pi\frac{p}{q} +2\alpha\right)}=-1$ as described in Eq.~(\ref{appen_eq:condition}).
The reason is as follows.
First, $e^{i2 \left[\alpha q +\frac{\pi}{2}(q+1) \right]} \neq -1$ indicates that $(\alpha+\pi/2)q=k \pi$ with $k$ being an integer.
Now that $p$ and $q$ are coprime, there should exist integers, $n_1$ and $n_2$, such that $n_1 p +n_2 q=1$ according to B\'{e}zout's identity,
which means, in turn, that any integer, say $k$, can be re-written as $(k n_1) p +(k n_2) q$.
The comparison between this condition and that in Eq.~(\ref{appen_eq:condition}) shows that they are in fact identical since one can always choose $l=k n_2$ and $m=-k n_1$.

Exponentiating the both sides of Eq.~(\ref{appen_eq:lnS2}) gives rise to the following result:
\begin{align}
S&= 1+ e^{i2 \left[\alpha q +\frac{\pi}{2}(q+1) \right]} \nonumber\\
&= 2 e^{i \left[\alpha q + \frac{\pi}{2}(q+1) \right]} \cos{\left( \alpha q + \frac{\pi}{2}(q+1) \right) } \nonumber\\
&= -2 e^{i \left[ \alpha q + \frac{\pi}{2}(q+1) \right]} \sin{\left( \left(\alpha +\frac{\pi}{2}\right) q\right)} .
\end{align}
By using this result, one can then show that Eq.~(\ref{appen_eq:left-hand}) becomes as follows:
\begin{align}
&\prod_{m=n+1}^{n+q} \left[ 2 \cos{\left(m\pi \frac{p}{q} +\alpha\right)}  \right] \nonumber \\
&= -2 e^{-i\frac{\pi}{2}p(2n+q+1) -i\alpha q} e^{i \left[\alpha q + \frac{\pi}{2}(q+1) \right]} \sin{\left(\left(\alpha +\frac{\pi}{2}\right)q\right)} \nonumber\\
&= 2 e^{-i\pi [ pn+1+(q+1)(p-1)/2 ]}  \sin{\left(\left(\alpha +\frac{\pi}{2}\right)q\right)}     \nonumber\\
&= 2 e^{i\pi\gamma_{pqn}}  \sin{\left(\left(\alpha +\frac{\pi}{2}\right)q\right)}  .
\label{appen_eq:final}
\end{align}
where $\gamma_{pqn}=pn+1+(q+1)(p-1)/2$.
Dividing the both sides of Eq.~(\ref{appen_eq:final}) by $2^q$ finally results in Eq.~(\ref{appen_eq:cosine_product}).

\section{Clausen approximation for the zero-energy mode}
\label{appendix:Clausen_derivation}

In this section of Appendix, we derive the analytic expression for the wave function profile of the zero-energy mode, which becomes exact in the weak-field limit, and provides a very approximation to the exact solution at moderately small flux values.
For completeness, here, we consider both the optimal and the Landau gauge.

In the case of the optimal gauge,
let us begin with the following Harper's equation for the zero-energy mode in sublattice B:
\begin{align}
\frac{\psi^\B_n}{\psi^\B_0}= \prod^{n}_{m=1} \left[  -A_m(\tilde{k}_y)   \right] ,
\label{appen_eq:ratio}
\end{align}
where $A_m(\tilde{k}_y)=2e^{i\left( m\pi\frac{\phi}{\phi_0} +\frac{\tilde{k}_y}{2}\right)} \cos{\left( m\pi\frac{\phi}{\phi_0} +\frac{\tilde{k}_y}{2} \right)}$.
Taking the absolute value and the logarithm of the both sides of Eq.~(\ref{appen_eq:ratio}) gives rise to the following:
\begin{align}
\ln{ \left| \frac{\psi^\B_{n}}{\psi^\B_{0}} \right|}
=\sum_{m=1}^n \ln{ \left[2 \left| \cos{ \left( m\pi\frac{\phi}{\phi_0}+\frac{\tilde{k}_y}{2} \right) } \right| \right] }
\label{appen_eq:ln_abs}
\end{align}

In the weak-field limit when $\phi/\phi_0 \ll 1$, one can approximate the summation in the right-hand side of Eq.~(\ref{appen_eq:ln_abs}) with an integral via the substitution of $x=m\pi\frac{\phi}{\phi_0}+\frac{\tilde{k}_y}{2}$ and $dx=\pi\frac{\phi}{\phi_0}$.
That is to say, by using the midpoint rectangle method, one can approximate the above summation as follows:
\begin{align}
\ln{\left|\frac{\psi^\B_{n}}{\psi^\B_{0}}\right|}
&\approx  \frac{1}{\pi\phi}\int_{x_1-\frac{\pi}{2}\frac{\phi}{\phi_0}}^{x_n+\frac{\pi}{2}\frac{\phi}{\phi_0}}    dx \ln{(2|\cos x|)} \nonumber \\
&=\frac{1}{\pi\phi}  \int_{x_1-\frac{\pi}{2}\frac{\phi}{\phi_0}}^{x_n+\frac{\pi}{2}\frac{\phi}{\phi_0}}   dx \ln{\left|1+e^{-2ix}\right|} \nonumber \\
&=\frac{1}{\pi\phi}\int_{x_1-\frac{\pi}{2}\frac{\phi}{\phi_0}}^{x_n+\frac{\pi}{2}\frac{\phi}{\phi_0}}  \sum_{s=1}^\infty (-1)^{s+1}\frac{\cos(2sx)}{s} \nonumber \\
&=-\frac{1}{\pi\phi} \sum_{s=1}^\infty\frac{\sin{ \left[ s(2x+\pi) \right] } }{2s^2} \Bigg|_{x_1-\frac{\pi}{2}\frac{\phi}{\phi_0}}^{x_n+\frac{\pi}{2}\frac{\phi}{\phi_0}} \nonumber \\
&=-\frac{1}{2\pi\phi}\Bigg[ \mathrm{Cl}_2\left(2x_n+\pi\frac{\phi}{\phi_0}+\pi\right)
\nonumber \\
&-\mathrm{Cl}_2\left(2x_1-\pi\frac{\phi}{\phi_0}+\pi\right) \Bigg] ,
\end{align}
where $x_n=n\pi\frac{\phi}{\phi_0} +\frac{\tilde{k}_y}{2}$ and $x_1=x_{n=1}$.
Note that $\textrm{Cl}_2 (\theta) = \sum^{\infty}_{n=1} \sin{(n\theta)}/n^2$ is called the Clausen function.
Neglecting the proportionality constant which is independent of $n$, we arrive at the final result:
\begin{align}
\left|\psi^\B_{n}\right| \propto \exp{\left[ -\frac{1}{2\pi\phi/\phi_0} \mathrm{Cl}_2 \left( 2\pi\frac{\phi}{\phi_0} n+\eta \right) \right]} ,
\end{align}
where $\eta=\tilde{k}_y+\pi(\phi/\phi_0+1)$.
By noting that Harper's equation for sublattice A is simply the inverse of that for sublattice B, one can obtain the following expression for the wave function profile in sublattice A:
\begin{align}
\left|\psi^\A_{n}\right| \propto \exp{\left[ \frac{1}{2\pi\phi/\phi_0} \mathrm{Cl}_2 \left( 2\pi\frac{\phi}{\phi_0} n+\eta \right) \right]} .
\end{align}

Now, let us switch gears to the Landau gauge, $\vec{A}=(0,\mathrm{B}x)$.
In the Landau gauge,
the hopping amplitude gains the following phase whose value is determined by the line integral between the nearest neighboring sites, $\phi_{ij}=\frac{e}{2\pi\hbar c}\int_i^j {\bf A} \cdot d {\bf l}$:
\begin{align}
\phi_{\alpha n \A,\alpha^\pr n^\pr \B} &=\left[ 2(\alpha^\pr-\alpha)-(-1)^n \right] \delta_{nn^\pr}\phi_n \nonumber \\
\phi_{\alpha n \B,\alpha^\pr n^\pr \A} &=\left[ 2(\alpha^\pr-\alpha)+(-1)^n \right] \delta_{nn^\pr}\phi_n
\end{align}
where $\phi_n=\frac{\phi}{\phi_0}\left(n/2-5/12\right)$.
As before, $n$ is the dimer index and $\alpha$ labels a unit cell along the $y$-direction (See Fig.~\ref{fig:optimal_gauge}).

At this point, it is convenient to consider a semi-infinite configuration of graphene with a zigzag edge, in which case the wave function amplitude on one of the sublattices can be chosen to be identically zero.
Defining sublattice B as the one with non-zero wave function amplitudes, one can show that the wave function amplitude in sublattice B is given as follows:
\begin{align}
\frac{\psi^{\B}_{n+1}}{\psi^\B_1} =\prod_{m=1}^n \left[ -2\cos{\left( m\pi\frac{\phi}{\phi_0} +\frac{\tilde{k}_y}{2}-\frac{5\pi}{6}\phi \right)} \right] .
\end{align}
Since the above formula is basically identical to that of the optimal gauge in Eq.~(\ref{appen_eq:ratio}), the same computation procedure previously applied in the optimal gauge can be performed to show that, in the weak-field limit,
\begin{align}
\left|\psi^\B_{n+1}\right| \propto \exp{\left[ -\frac{1}{2\pi\phi/\phi_0} \mathrm{Cl}_2 \left( 2\pi\frac{\phi}{\phi_0} n+\kappa \right) \right]} ,
\end{align}
where $\kappa=\tilde{k}_y -\frac{5\pi}{3}\phi/\phi_0+\pi$.




\end{document}